\documentclass{article} 
\usepackage{colm2024_conference}

\makeatletter
\fancyhfoffset[L]{0pt}
\fancyhfoffset[R]{0pt}
\makeatother

\usepackage{microtype,multirow}
\usepackage{booktabs,enumitem}
\usepackage{url}
\usepackage{amsmath,amssymb}
\usepackage{caption}
\usepackage{float}
\usepackage{placeins}
\usepackage{xcolor} 
\usepackage{graphicx}
\usepackage{tikz}
\usepackage{hyperref} 
\usepackage{fancyhdr}
\usepackage{setspace}
\makeatletter
\providecommand{\github}{}
\providecommand{\huggingface}{}
\providecommand{\modelscope}{}
\providecommand{\ghlink}{}
\providecommand{\hflink}{}
\providecommand{\mslink}{}
\makeatother

\newcommand{\firstpagelogobox}{\includegraphics[height=6mm]{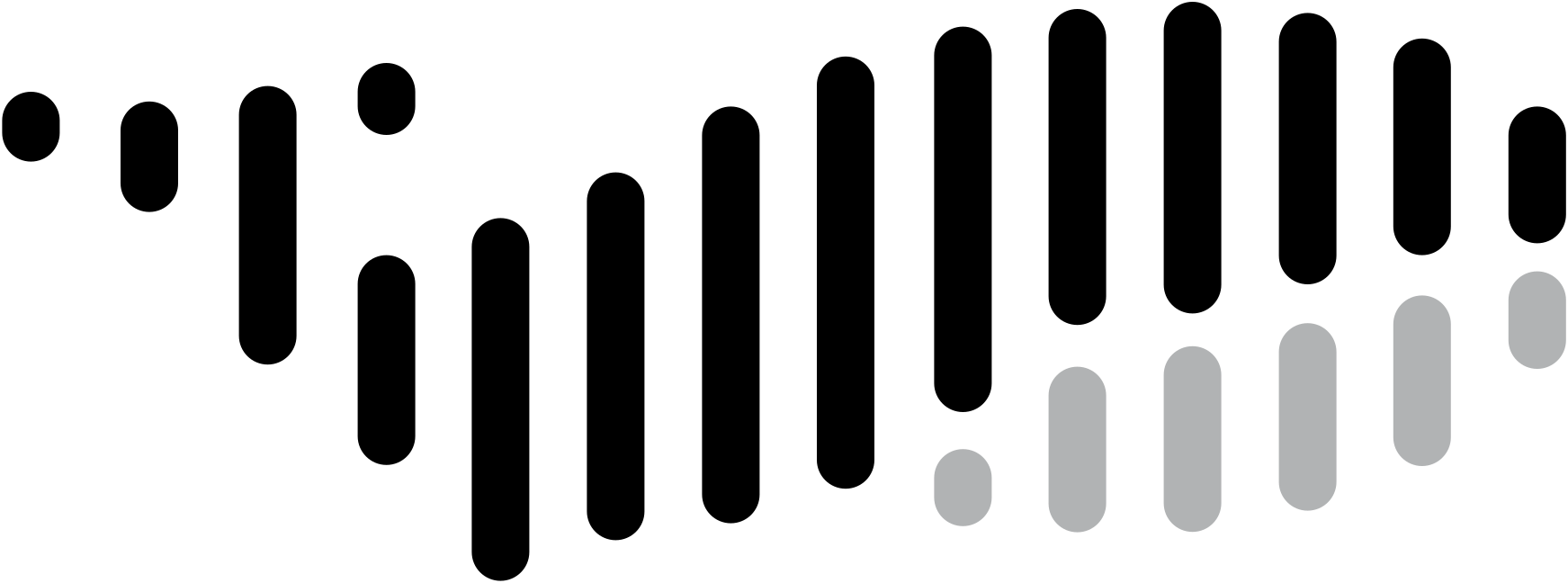}}
\definecolor{darkblue}{rgb}{0, 0, 0.5}
\hypersetup{colorlinks=true, citecolor=darkblue, linkcolor=darkblue, urlcolor=darkblue}

\title{Fish Audio S2 Technical Report}
\date{} 

\author{\centering Fish Audio Team\thanks{Please send correspondence to opensource@fish.audio.}}
\makeatletter
\def\@author{\vspace{-0.4em}\centering Fish Audio Team\vspace{-0.8em}}
\makeatother

\begin{document}

\pagestyle{fancy}
\fancyhf{}
\fancyhead{}
\renewcommand{\headrulewidth}{0.8pt}
\renewcommand{\footrulewidth}{0.8pt}
\fancyhfoffset[L]{8pt}
\fancyhfoffset[R]{0pt}
\fancyfoot[L]{}
\fancyfoot[C]{\footnotesize \strut\thepage}
\renewcommand{\footrulewidth}{0pt}

\setlength{\headheight}{30pt}
\fancypagestyle{firstpage}{%
  \fancyhf{}
  \fancyhead[L]{\raisebox{8pt}{\firstpagelogobox\hspace{6pt}\raisebox{3.5pt}{\includegraphics[height=3mm]{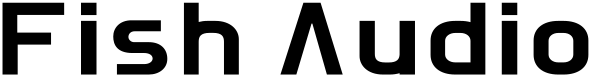}}}}
  \renewcommand{\headrulewidth}{0.6pt}
  \renewcommand{\footrulewidth}{0.6pt}
  \setlength{\headwidth}{\textwidth}
  \fancyhfoffset[L]{0pt}
  \fancyhfoffset[R]{0pt}
  \fancyfoot[L]{\footnotesize \strut\copyright\ 2026 39 AI, Inc All rights reserved.}
  \fancyfoot[C]{}
}
\setlength{\footskip}{24pt}

\vspace*{-0.3cm}
\maketitle
\thispagestyle{firstpage}
\fancyhead[L]{}
\fancyhead[C]{}
\fancyhead[R]{}

\vspace{-40pt}
\begin{center}
  {\small
    \href{https://github.com/fishaudio/fish-speech}{\raisebox{-0.2ex}{\includegraphics[height=1.7ex]{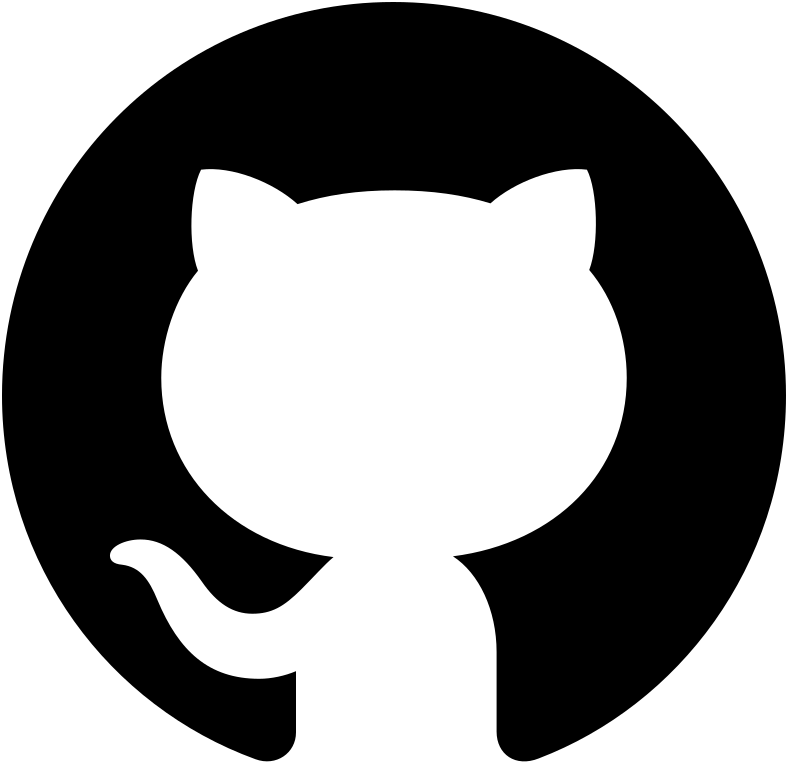}}}\quad\url{https://github.com/fishaudio/fish-speech}\par
    \href{https://huggingface.co/fishaudio/s2-pro}{\raisebox{-0.2ex}{\includegraphics[height=1.7ex]{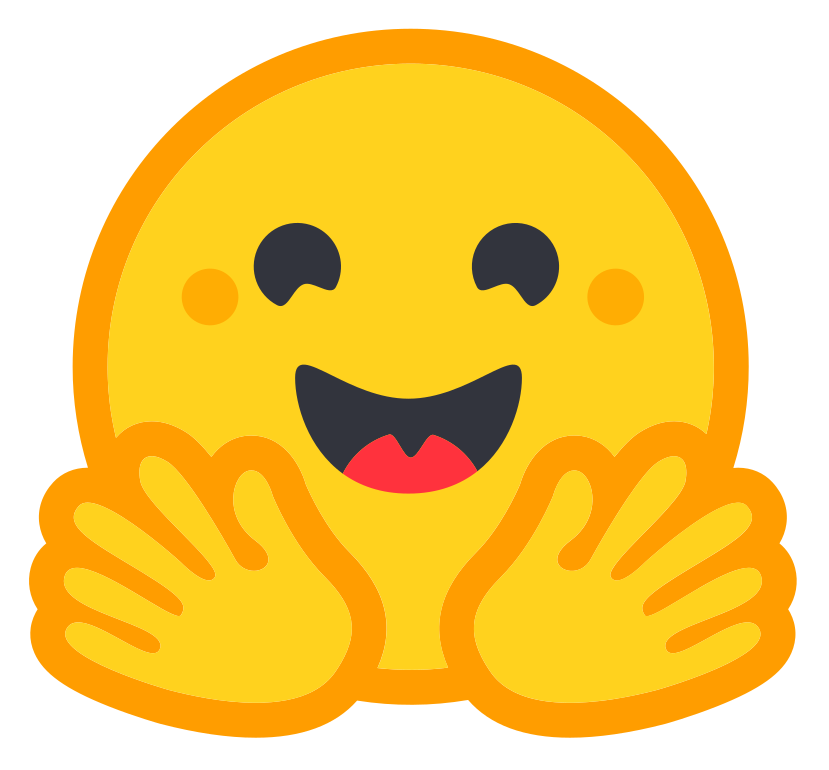}}}\quad\url{https://huggingface.co/fishaudio/s2-pro}\par
    {\raisebox{-0.2ex}{\includegraphics[height=1.7ex]{logos/logo.png}}}\quad\url{https://fish.audio}\par
  }
\end{center}

\let\oldabstract\abstract
\let\endoldabstract\endabstract
\renewenvironment{abstract}{
  \begin{center}
    \begin{minipage}{0.88\textwidth}
    \begin{oldabstract}}{
      \end{oldabstract}
    \end{minipage}
\end{center}}

\begin{abstract}
  \vspace{10pt}

  We introduce Fish Audio S2, an open-sourced text-to-speech system featuring multi-speaker, multi-turn generation, and, most importantly, instruction-following control via natural-language descriptions.
  To scale training, we develop a multi-stage training recipe together with a staged data pipeline covering video captioning and speech captioning, voice-quality assessment, and reward modeling.
  To push the frontier of open-source TTS, we release our model weights, fine-tuning code, and an SGLang-based inference engine.
  The inference engine is production-ready for streaming, achieving an RTF of 0.195 and a time-to-first-audio below $100\,\text{ms}$.
  Our code and weights are available on \textbf{\href{https://github.com/fishaudio/fish-speech}{GitHub}} and \textbf{\href{https://huggingface.co/fishaudio/s2-pro}{Hugging Face}}.
  We highly encourage readers to visit \textbf{\url{https://fish.audio}} to try custom voices.

\end{abstract}

\section{Introduction}

\begin{figure}[!htbp]
  \centering
  \includegraphics[width=0.9\linewidth]{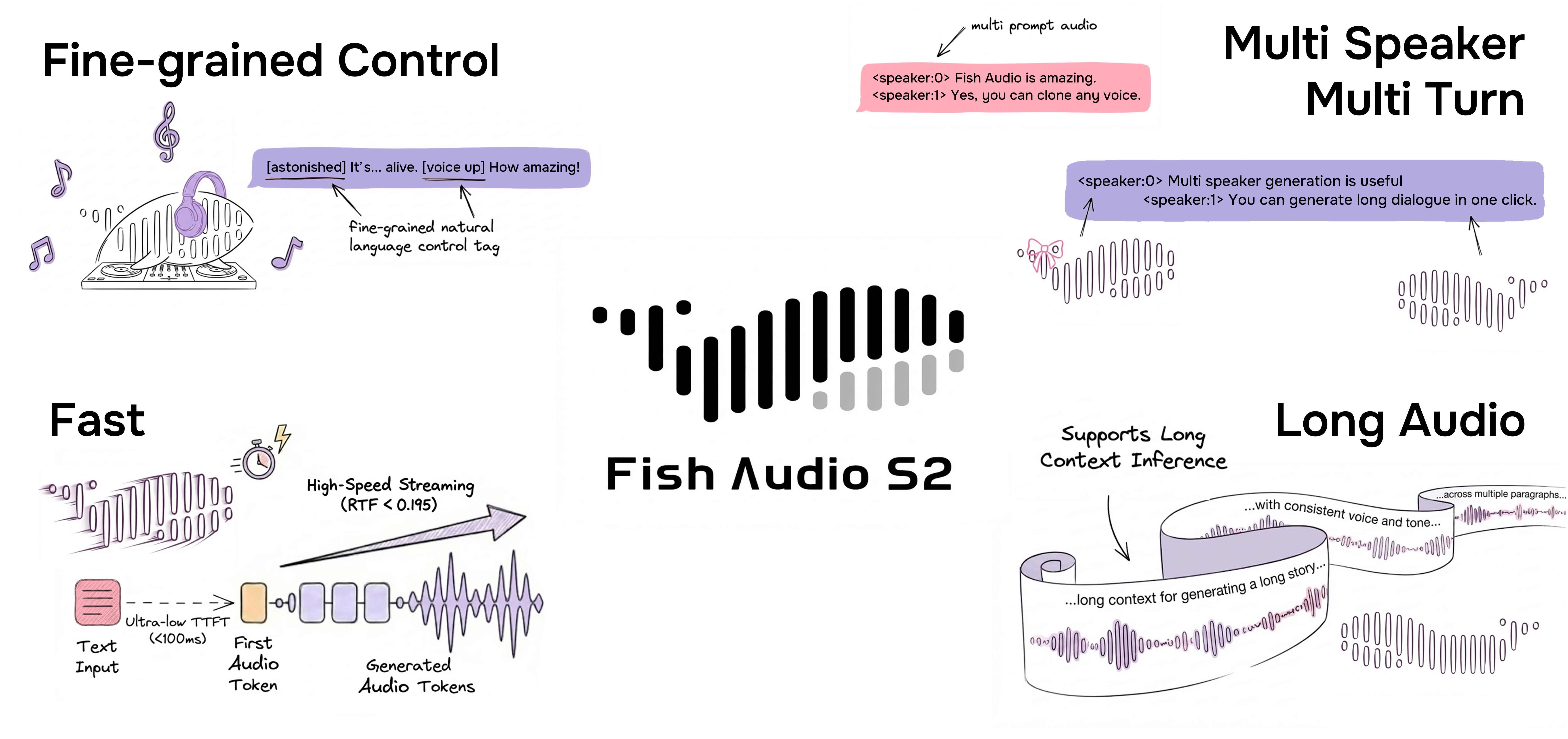}
  \caption{\textbf{Fish Audio S2 is a multilingual, controllable, and expressive TTS system supporting long-form, multi-speaker, multi-turn generation with ultra-low TTFA and RTF.}}
  \label{fig:placeholder-feature}
\end{figure}

High-quality, controllable text-to-speech (TTS) has become increasingly important in modern AI systems, enabling scalable audio content creation and natural conversational experiences across applications such as audiobook narration, video dubbing, and personalized chatbots.
Recent progress in TTS has been driven by large-scale models \citep{zhang2025minimaxspeechintrinsiczeroshottexttospeech,du2025cosyvoice3inthewildspeech,li2026indextts25technicalreport,hu2026qwen3ttstechnicalreport}.
Many of these works follow a two-stage paradigm: conditioned on text, the model first produces high-level discrete speech tokens, which are then decoded into the full waveform by a separate acoustic decoder \citep{wang2023neuralcodeclanguagemodels, défossez2022highfidelityneuralaudio,kong2020hifi,anastassiou2024seedttsfamilyhighqualityversatile}.

Alongside these architectural innovations, the success of large-scale TTS relies heavily on robust data curation.
Recent efforts have introduced sophisticated pipelines for cleaning speech corpora and annotating paralinguistic features~\citep{Cheng_2025,yang2025gigaspeech}. However, generating fine-grained natural-language instructions for vocal features at scale remains a major bottleneck.
From a training perspective, although reinforcement learning (RL) methods such as Direct Preference Optimization (DPO)~\citep{rafailov2023direct}, Proximal Policy Optimization (PPO)~\citep{schulman2017proximal} and Group Relative Policy Optimization (GRPO)~\citep{shao2024deepseekmath} have become standard for improving model behavior in the large language model (LLM) domain~\citep{guo2025deepseek,agarwal2025gpt}, their adoption in TTS remains limited.

In this report, we present Fish Audio S2, which retains the decoder-only Transformer backbone and RVQ-based audio codec of Fish Audio S1~\citep{liao2024fishspeechleveraginglargelanguage}. We extends it with a unified data curation and RL alignment framework to improve controllability, naturalness, and robustness in speech generation.
Specifically, we introduce two key technical innovations:
\begin{itemize}[leftmargin=2em, itemsep=-0.3em, topsep=-0.2em]
  \item \textbf{Multi-Purpose Data Pipeline.}
    We build a data pipeline with a speech quality assessment model and a rich-transcription ASR model to filter and annotate large-scale audio data for TTS pre-training. The same models are then directly reused as reward signals for RL alignment, eliminating distribution mismatch between the two stages.
  \item \textbf{Multi-Reward RL Alignment.} We implement a variant of GRPO that jointly optimizes semantic accuracy, acoustic quality, and speaker similarity, ensuring a balance between expressiveness and robustness.
\end{itemize}

These innovations directly enable three major functional breakthroughs:
\begin{itemize}[leftmargin=2em, itemsep=-0.3em, topsep=-0.2em]
  \item \textbf{Enhanced Instruction Following.} Fish~Audio~S2 exhibits superior adherence to natural language instructions. It enables broad and fine-grained control over speech generation through free-form textual descriptions.
  \item \textbf{Native Multi-Speaker and Multi-Turn Generation.} The model can natively generate complex, interleaved dialogues involving multiple distinct speakers in a single pass, capturing the dynamics of natural conversation.
  \item \textbf{Stable Long-Form Synthesis.} The system supports the generation of coherent and continuous audio, maintaining stability and consistency over extended durations.
\end{itemize}

To evaluate our model, we conduct extensive experiments along two complementary tracks: (i) objective evaluation and (ii) LLM-as-a-Judge assessments of higher-level capabilities.
For intelligibility, content accuracy, long-form, and multilingual performance, we report Word Error Rate (WER), Character Error Rate (CER), and speaker similarity on widely used benchmarks, including Seed-TTS-Eval~\citep{anastassiou2024seedttsfamilyhighqualityversatile}, MiniMax~Multilingual~Testset~\citep{zhang2025minimaxspeechintrinsiczeroshottexttospeech}, CosyVoice3-Eval~\citep{du2025cosyvoice3inthewildspeech}, Long-TTS-Eval~\citep{wang2025mgm}.
Across public benchmarks, Fish~Audio~S2 shows consistently strong objective performance, achieving leading results on Seed-TTS benchmark while maintaining robust multilingual intelligibility and speaker similarity on both the MiniMax Multilingual Testset and CV3-Eval.
To assess higher-level capabilities such as instruction following and human-likeness, we further employ the Audio~Turing~Test~\citep{wang2025att} and Emergent~TTS~Eval~\citep{manku2025emergenttts}.
On the Audio~Turing~Test, Fish~Audio~S2 achieves a posterior mean of 0.483, which further improves to 0.515 with instruction rewriting. On Emergent~TTS~Eval, it reaches an overall win rate of 81.88\% against the baseline, further supporting its strong instruction-following capability.
Furthermore, to address the lack of dedicated benchmarks for fine-grained control, we introduce a novel evaluation benchmark, the Fish Audio Instruction Benchmark, which systematically evaluates models' inline tag-following performance across English and Chinese.
On the Fish Audio Instruction Benchmark, Fish Audio S2 achieves an overall tag-activation rate of 93.3\% and an overall quality score of 4.51/5.0 across English and Chinese, as evaluated by Gemini 3 Pro.

To accelerate research and lower the barrier to high-quality TTS development, we publicly release our model weights, fine-tuning code, and the SGLang-based inference engine on \textbf{\href{https://github.com/fishaudio/fish-speech}{GitHub}} and \textbf{\href{https://huggingface.co/fishaudio/s2-pro}{Hugging Face}}.
We also highly encourage readers to explore interactive demos at our official site \textbf{\url{https://fish.audio/}}.

The remainder of this paper is organized as follows: Section~\ref{sec:architecture} details the model architecture; Section~\ref{sec:data} describes the data curation pipeline; Section~\ref{sec:training} presents the pre-training and RL-based post-training; Section ~\ref{sec:inference} instrodces our inference engine and its performance, Section~\ref{sec:evaluation} presents the experimental setup and comprehensive evaluation results; and finally, Section~\ref{sec:conclusion} concludes with a discussion on limitations and future directions.

\section{Architecture}\label{sec:architecture}

\thispagestyle{fancy}
\begin{figure}[t]
  \centering
  \includegraphics[width=0.8\linewidth]{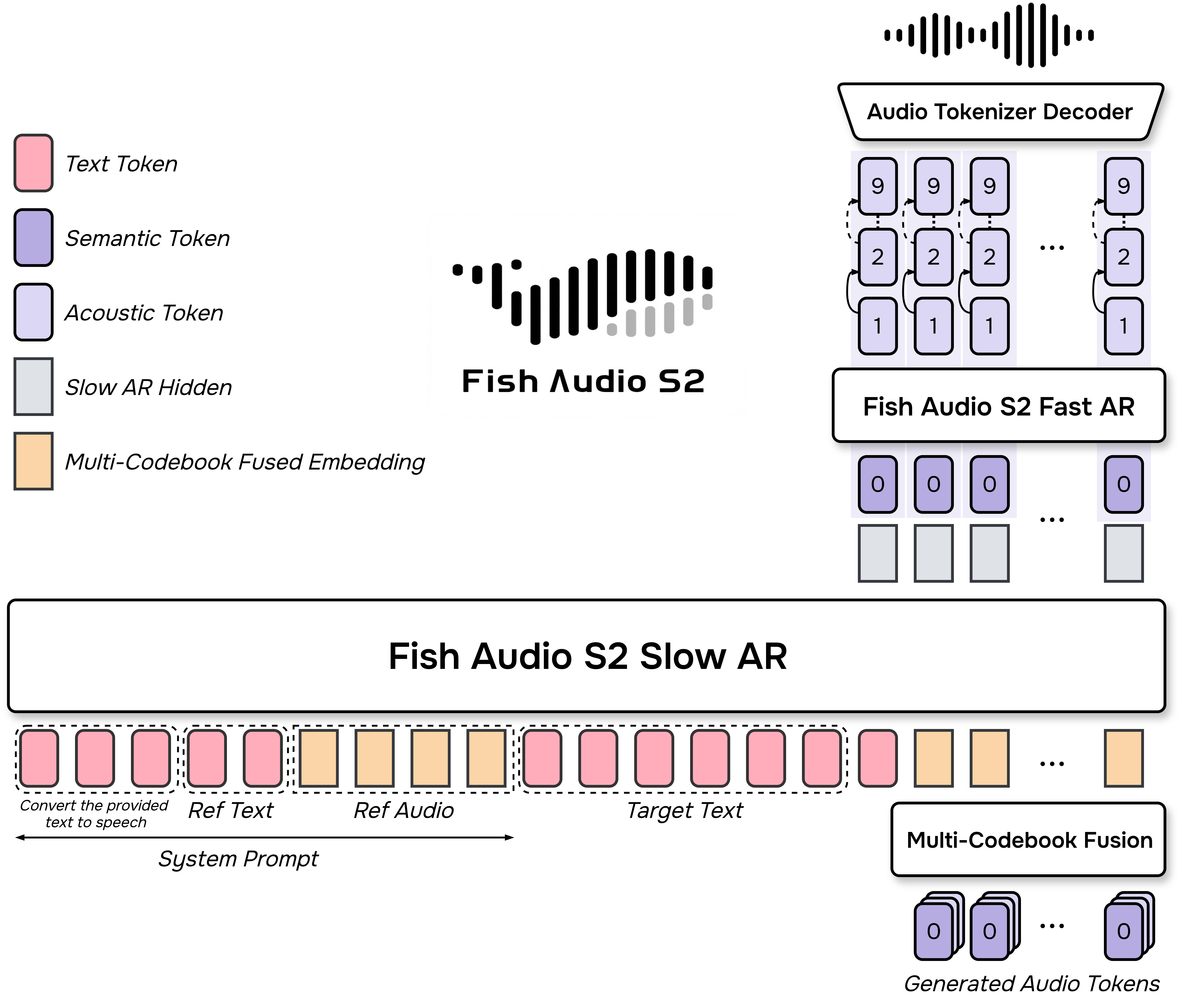}
  \caption{\textbf{Fish~Audio~S2 architecture.}}
  \label{fig:placeholder-Architecture}
\end{figure}

\subsection{Audio Tokenizer}

Our audio tokenizer is built upon the architecture of the Descript Audio Codec (DAC)~\citep{kumar2023high}, optimized for high-fidelity, real-time streaming at a $44.1\,\text{kHz}$ sampling rate.
The model employs a hierarchical Residual Vector Quantization (RVQ) strategy utilizing $N$ codebooks ($N{=}10$ in our model): the primary codebook serves as the semantic codebook, while the remaining nine capture progressively finer-grained acoustic details.

\textbf{Streaming Architecture.}
To adapt the vanilla DAC for low-latency TTS tasks, we introduce several key modifications to the encoder and decoder structures:
\begin{itemize}[leftmargin=2em, itemsep=-0.3em, topsep=-0.2em]
  \item \textbf{Causal Convolutions.} We refactor the model to be strictly causal by replacing standard convolutions with masked causal convolutions.
    This ensures the generation process depends solely on past context, enabling low-latency streaming capabilities.

  \item \textbf{Transformer Bottleneck.}
    Following the design of Mimi~\citep{defossez2024moshi}, we integrate causal sliding-window Transformer blocks both before and after the RVQ layers.
    By restricting attention to a fixed-size window, this mechanism models long-range dependencies with bounded memory usage, preventing out-of-memory issues during long-form inference.

  \item \textbf{Extended Downsampling.}.
    The encoder extends the standard DAC encoder ($512\times$) with additional ConvNeXt~V2~\citep{woo2023convnext} layers ($4\times$), achieving a total downsampling ratio of 2048 and a compact frame rate of approximately $21\,\text{Hz}$.

  \item \textbf{EVA-GAN Decoder.}
    Instead of the original DAC decoder, we employ the structure of EVA-GAN~\citep{liao2024eva} as our generator, which significantly improves parameter efficiency and synthesis quality, providing a more robust reconstruction of fine-grained acoustic details compared to the original DAC decoder.
\end{itemize}

\textbf{Semantic Distillation.}
To ensure that the first codebook captures rich linguistic and phonetic information, we adopt semantic distillation following~\citep{defossez2024moshi}.
During training, an auxiliary semantic prediction head is jointly optimized to regress the 16th-layer activations of a pre-trained w2v-BERT~2.0 model~\citep{barrault2023seamless}.
By feeding the quantized features from the first codebook into this head, we encourage the bottleneck to retain rich semantic representations, thereby enabling more stable alignment in downstream TTS.

\subsection{Dual-Autoregressive Generation}

When modeling high-fidelity acoustic features extracted by the audio tokenizer, directly flattening the 10-layer RVQ codebooks along the time axis leads to a tenfold increase in sequence length, severely limiting the LLM's ability to handle long contexts.
To address this dimensionality challenge, we apply a Dual-Autoregressive (Dual-AR) architecture (\cite{liao2024fishspeechleveraginglargelanguage}) that decouples temporal semantic modeling from depth-wise acoustic modeling, as illustrated in Figure~\ref{fig:placeholder-Architecture}.
This architecture comprises a core Temporal Semantic Backbone (Slow AR) coupled with a lightweight Depth-wise Acoustic Decoder (Fast AR).

\textbf{Slow AR.}
We adopt a pretrained Qwen3-4B as the Slow AR backbone.
The Slow AR operates autoregressively over the full token sequence, which interleaves text tokens (e.g. system prompts, target text) with discrete audio tokens.
During audio generation, it predicts the semantic token $q^{(0)}_t$ from the first RVQ codebook at each time step $t$.
Since this codebook undergoes semantic distillation during tokenizer training, the Slow AR can effectively plan linguistic content and coarse prosodic structure, analogous to standard text generation.

\textbf{Fast AR.}
Given the semantic tokens generated by the Slow AR, we introduce a lightweight Fast AR network---consisting of 4 Transformer layers with independent weights and embedding tables---to reconstruct the remaining fine-grained acoustic details.
At each time step $t$, the Slow AR first samples the semantic token $q^{(0)}_t$ and emits a hidden state $\mathbf{h}_t^{\text{slow}}$.
The Fast AR then generates the remaining $N{-}1$ acoustic tokens $q^{(1)}_t, \dots, q^{(N-1)}_t$ through a depth-wise autoregressive process.
The hidden state $\mathbf{h}_t^{\text{slow}}$ is first linearly projected to the Fast AR's dimension and placed at position~0 as a conditioning prefix, providing global context from the Slow AR.
The semantic token $q^{(0)}_t$, already determined by the Slow AR, is then embedded and placed at position~1 as the seed input.
The Fast AR then autoregressively generates $q^{(1)}_t$ through $q^{(N-1)}_t$, where each step conditions on the conditioning prefix $\mathbf{h}_t^{\text{slow}}$ and all previously generated tokens.
All $N$ codebook layers share a single embedding table within the Fast AR; the codebook layer identity is encoded through RoPE positional embeddings.
This highly asymmetric design---a 4B-parameter model along the time axis and a 4-layer network along the codebook depth axis---ensures high inference efficiency.

\textbf{Multi-Codebook Fusion (MCF).}
After all $N$ codebook tokens for time step $t$ have been generated, they are aggregated into a single continuous vector $\mathbf{x}_{t+1}$ to serve as the Slow AR's input embedding for the next time step $t+1$.
Each token $q^{(k)}_t$ ($k \in \{0, 1, \dots, N - 1\}$) is embedded via a dedicated embedding layer $\mathbf{E}^{(k)}$ that maps codebook indices into the Slow AR's embedding space.
These $N$ codebook embeddings, together with the Slow AR's own token embedding $\mathbf{e}^{\text{LM}}_t$ for the semantic token $q^{(0)}_t$, are summed:
\begin{equation}
  \mathbf{x}_{t+1} = \mathbf{e}^{\text{LM}}_t + \sum_{k=0}^{N-1} \mathbf{E}^{(k)}\bigl[q^{(k)}_t\bigr],
\end{equation}
where $N=10$ is the total number of codebooks.
Note that the semantic token $q^{(0)}_t$ contributes two distinct representations: $\mathbf{e}^{\text{LM}}_t$ from the Slow AR's token embedding layer, and $\mathbf{E}^{(0)}[q^{(0)}_t]$ from the codebook embedding layer.
These two embedding tables are independently parameterized and capture complementary aspects of the same token.

\section{Data Pipeline}\label{sec:data}

\thispagestyle{fancy}
\begin{figure}[t]
  \centering
  \includegraphics[width=\linewidth]{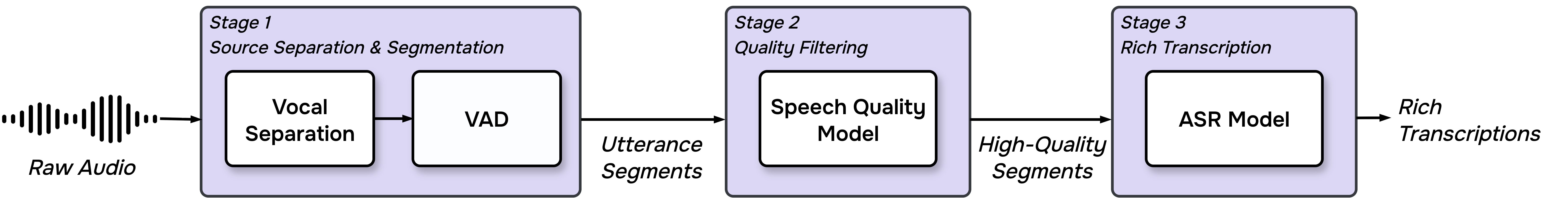}
  \vspace{1pt}
  \caption{\textbf{Fish~Audio~S2 data pipeline.}}
  \label{fig:data_pipeline}
\end{figure}

Scaling TTS systems requires massive, high-quality datasets.
Beyond basic noise reduction, the primary bottleneck lies in mapping subtle acoustic attributes (e.g., emotion and prosody) and speaker turns to natural language instructions---a process that is infeasible to scale manually.
Moreover, RL alignment for TTS typically relies on reward models trained independently from the pre-training pipeline, which can introduce distribution shift between pre-training data and post-training objectives.

To address both challenges, we design a three-stage data curation pipeline built around two core evaluation engines: a speech quality model and a rich-transcription ASR model.
During pre-training, these engines act as strict filters and annotators; during RL-based post-training, they are directly reused as reward models.
This dual-purpose design eliminates distribution shift between pre-training and post-training by construction, while enabling fine-grained vocal annotation in natural language to scale automatically without human intervention.

To process raw audio into speech-text pairs with fine-grained vocal annotations, our pipeline executes three stages (Figure~\ref{fig:data_pipeline}):
\begin{itemize}[leftmargin=2em, itemsep=0.1em, topsep=0.2em]
  \item \textbf{Stage 1: Source Separation and Segmentation.} We apply a vocal separation module to isolate clean speech from background noise, followed by Voice Activity Detection (VAD) to slice continuous audio into utterance-level segments.

  \item \textbf{Stage 2: Quality Filtering.} Our core speech quality model evaluates each utterance across multiple dimensions---including signal-to-noise ratio, speaker consistency, recording quality, and intelligibility---to filter out low-fidelity samples.
  
  \item \textbf{Stage 3: Rich Transcription.} An in-house ASR model generates highly accurate transcripts.
  This model simultaneously transcribes long-form spoken text and captions vocal features (e.g., emotion, prosody, paralinguistic) and speaker turns, creating descriptive natural language captions that directly enable the model's zero-shot instruction-following capabilities.
\end{itemize}

\subsection{Speech Quality Model}
Following the architectural design of Uni-VERSA~\citep{shi2025uni}, our speech quality model utilizes a pre-trained w2v-BERT~2.0 backbone coupled with a multi-layer perceptron head for acoustic evaluation.
We train this network on a proprietary dataset of thousands of hours of Stage~1 audio with speech quality labels provided by human annotators, using a combined objective of MSE and focal loss~\citep{lin2017focal}.
In Stage~2, this model acts as a strict filter, removing low-quality samples that slip through Stage~1, such as overlapping voices and residual background music, significantly reducing artifacts such as timbre inconsistency in the pre-training data.
Consistent with our dual-purpose design, this same model is reused during the RL phase as an objective acoustic reward, penalizing noise and artifacts in the generated speech.

\subsection{Rich-Transcription ASR Model}

We develop a rich-transcription ASR model by fine-tuning the Qwen3-Omni-30B-A3B foundation model to jointly transcribe spoken content and annotate speaker turns and vocal events.
The training data were curated using a video-based pseudo-labeling approach, followed by human verification to ensure annotation accuracy.

\begin{figure}[t]
  \centering
  \includegraphics[width=0.8\linewidth]{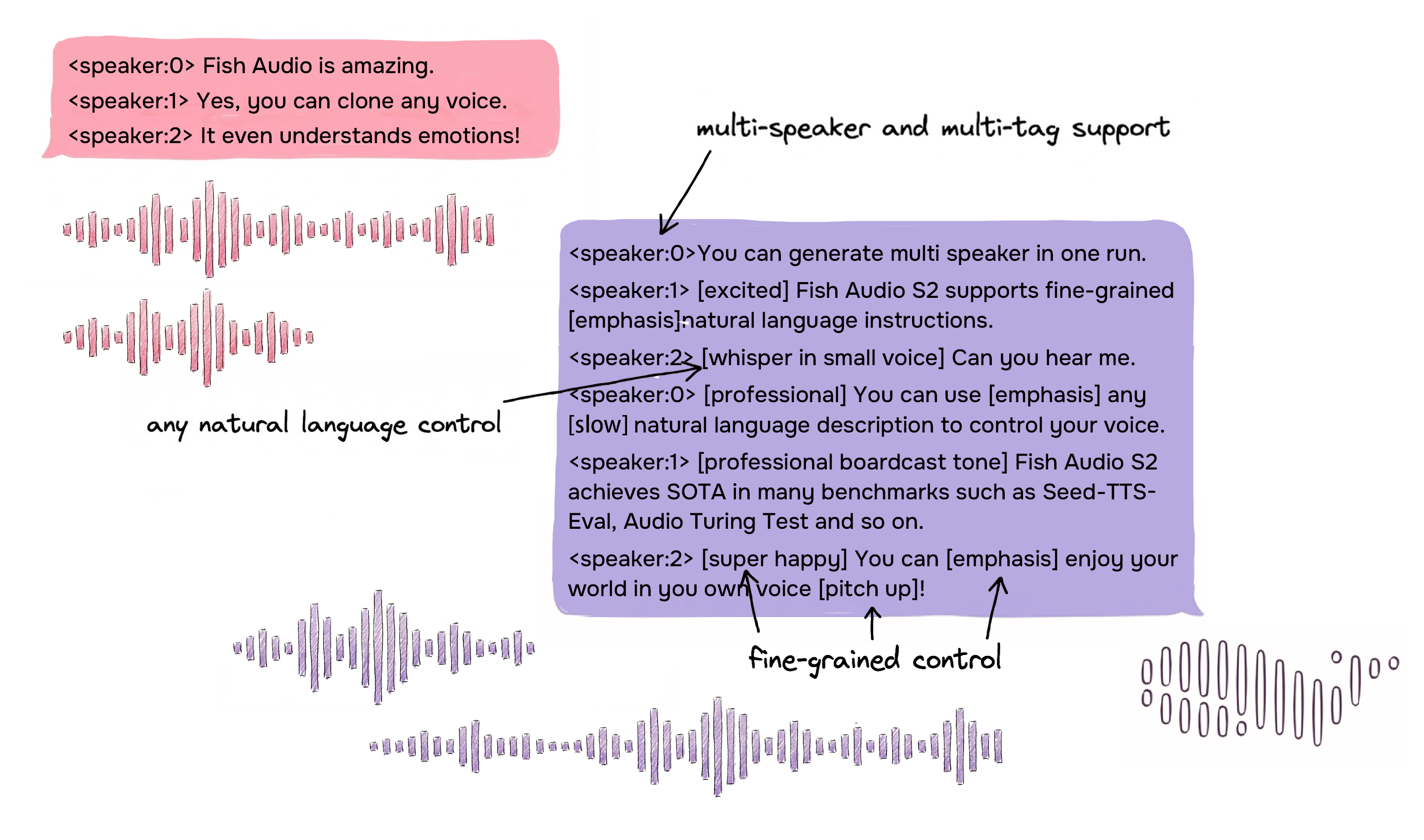}
  \caption{\textbf{Fish Audio S2 supports multi-speaker generation with fine-grained natural language control over prosody, emotion, and speaking style.}}
  \label{fig:test_sample}
\end{figure}

In Stage~3, this model jointly transcribes spoken content and annotates vocal features as natural language instructions.
Specifically, it predicts speaker turns (e.g., \texttt{<|speaker:0|>}) and injects vocal instructions such as \texttt{[prolonged laugh]}, \texttt{[inhale]}, \texttt{[angry]}, \texttt{[emphasis]} and \texttt{[in a hurry]} directly into the text stream alongside natural disfluencies.
An example of the output format is shown in Figure~\ref{fig:test_sample}.
These transcripts serve as the fine-grained natural language instructions for training the zero-shot controllable generation capabilities of Fish~Audio~S2.

Consistent with our dual-purpose design, this model is reused during RL-based post-training as an intelligibility and instruction-following reward.
By re-transcribing the generated audio and comparing it against the original prompt, it provides reward signals that penalize hallucinations, missing words, and ignored vocal instructions.

\section{Training}\label{sec:training}

The training pipeline of Fish~Audio~S2 proceeds in four stages.
We first train the audio tokenizer to obtain discrete audio representations, then progressively align the LLM with these discrete representations through large-scale pre-training and supervised fine-tuning (SFT) on curated data, and finally refine generation quality via RL-based post-training.

\subsection{Audio Tokenizer Training}

The complete audio tokenizer, totaling 446M parameters, is trained for ~1M steps.
We employ a composite GAN loss framework to ensure perceptual fidelity, utilizing three distinct discriminators: a multi-period discriminator to capture periodic signals, a multi-resolution discriminator for spectral consistency, and a multi-scale STFT discriminator to ensure high-frequency detail and phase coherence.

\subsection{Pre-training and SFT}
\label{sec:pretrain}

The pre-training phase aligns the audio tokens with the Qwen3-4B foundation model through two progressive stages.
The first stage establishes cross-modal alignment with a maximum context length of 8,192 tokens; the second stage extends the context to 16,384 tokens, enabling long-form audio synthesis and multi-turn multi-speaker conversational generation.
In total, the pre-training phase utilizes over 10 million hours of raw audio across approximately 80 languages and dialects.
After pre-training, we perform SFT on curated internal high-quality labelled data to improve expressiveness and controllability.

We expanded the Qwen3-4B vocabulary with structural control tokens and 4,096 semantic tokens.
To ensure a smooth feature space transition, the new token embeddings are initialized by sampling from a multivariate normal distribution $\mathcal{N}(\mu, \Sigma)$, where $\mu$ and $\Sigma$ are the empirical mean and covariance of the existing text embedding matrix.

Unlike Fish Audio S1, which appends reference audio to the user input, S2 prepends the reference audio to the system prompt.
The loss of the reference audio tokens is masked during training to prevent verbatim memorization.
For fine-grained acoustic control, rather than relying on lengthy global prompts, we inject descriptive instructions at specific word or phrase positions within the dialogue context, enabling precise localized control over acoustic details.
These instructions take the form of natural language---such as \texttt{whisper}, \texttt{angry}, and \texttt{laugh}---embedded directly in the token sequence.
Through autoregressive training on large-scale data, the model naturally internalizes the mapping between these textual cues and localized acoustic variations without requiring dedicated control tokens.

The training objective follows standard autoregressive language modeling, adapted for the Dual-AR architecture with a separate loss for each component.
For the Slow AR, the training objective is defined as:
\begin{equation}
  \mathcal{L}_{\text{slow}} = - \sum_{t=0}^{T-1} m_t \, \lambda_t \, \log P(x_t \mid x_{<t}),
\end{equation}
where $m_t \in \{0,1\}$ is the reference mask ($m_t = 0$ for system prompt and reference audio tokens, $m_t = 1$ otherwise).
The Fast AR loss $\mathcal{L}_{\mathrm{fast}}$ supervises the depth-wise generation of audio tokens $q^{(0)}_t, \dots, q^{(N-1)}_t$, conditioned on the Slow AR hidden state $\mathbf{h}_t^{\text{slow}}$:
\begin{equation}
  \mathcal{L}_{\text{fast}} = - \frac{1}{\sum_1^{N-1}w^{(k)}} \sum_{k=0}^{N-1} w^{(k)} \, \log P(q_t^{(k)} \mid \mathbf{h}_t^{\text{slow}}, q_t^{(<k)}),
\end{equation}
where $w^{(k)}$ is the weight for the $k$-th acoutic token.
During pre-training, $w^{(k)} = 1$ uniformly for all codebook layers including $k=0$, where predicting the semantic token serves as an auxiliary objective that helps the Fast AR learn to extract information from the Slow AR's projected hidden state.
During SFT, we remove the semantic token prediction and apply a progressively decayed weighting strategy over the remaining codebooks to better align training with the inference setting, where the semantic token $q_t^{(0)}$ is sampled from the Slow AR, while concentrating model capacity on the coarse-grained acoustic codebooks that contribute most to perceptual quality.
The total loss combines both components:
\begin{equation}
  \mathcal{L}_{\text{total}} = \lambda_{\text{slow}} \,\mathcal{L}_{\text{slow}} + \lambda_{\text{fast}} \,\mathcal{L}_{\text{fast}}.
\end{equation}

The entire pre-training framework is built upon Fully Sharded Data Parallel (FSDP), with a differential learning rate strategy that applies a reduced learning rate to the text foundation parameters while using a higher learning rate for the audio modules.
Combined with a Warmup-Stable-Decay (WSD) scheduling strategy~\citep{hu2024minicpm}, this ensures stable training at scale with high throughput.

\subsection{RL-Based Post-Training}

Following the pre-training and SFT phases, the RL-based post-training phase aims to mitigate hallucinations, token skipping, and timbre drift commonly observed in autoregressive audio generation.
Audio generation involves extremely long sequences, making standard PPO prohibitively expensive due to the need to maintain a large value model in memory.
We therefore adopt an RL algorithm inspired by GRPO~\citep{shao2024deepseekmath} and Dr.GRPO~\citep{liu2025understanding}, which eliminates the value network entirely by estimating advantages from group-level statistics.
Specifically, for a given prompt, the model independently samples $G$ candidate outputs $\{y_1, \dots, y_G\}$ and computes advantages:

\begin{equation}
  A_i = \mathcal{R}_i - \bar{\mathcal{R}}, \quad i \in \{1, \dots, G\},
\end{equation}

where $\mathcal{R}_i$ is the reward for the $i$-th candidate, $\bar{\mathcal{R}}$ is the intra-group mean.
Following Dr.GRPO, we remove normalization by the intra-group standard deviation to avoid sample-level difficulty bias, where samples with low reward variance receive disproportionately large gradient updates.
These advantages are then used to optimize both components of the Dual-AR architecture.
The Slow AR policy loss is defined as:
\begin{equation}
  \mathcal{L}_{\text{slow}}^{\text{RL}} = -\frac{1}{T} \sum_{t=1}^{|T|} A_i \, \log \pi_\theta(x_t \mid x_{<t}) + \beta \, D_{\mathrm{KL}}^{(t)},
\end{equation}

where $D_{\mathrm{KL}}^{(t)}$ is the per-token KL divergence between the current policy and the reference policy, computed via the Schulman estimator.
The Fast AR loss follows the same formulation but operates independently over each audio token $q_t^{(k)}$, sharing the same advantage signal:
\begin{equation}
  \mathcal{L}_{\text{fast}}^{\text{RL}} = -\frac{1}{\mathcal{C}^{(k)}} \sum_{t,k} A_i \, \log \pi_\theta^{\mathrm{FA}}(q_t^{(k)} \mid q_t^{(<k)}) + \beta \, D_{\mathrm{KL}}^{(t,k)},
\end{equation}
where $\mathcal{C}^{(k)}$ is the $k$-th codebook size.
The total RL loss combines both components:
\begin{equation}
  \mathcal{L}_{\text{RL}} = \mathcal{L}_{\text{slow}}^{\text{RL}} + \gamma \, \mathcal{L}_{\text{fast}}^{\text{RL}}.
\end{equation}

As speech generation quality spans multiple perceptual dimensions, we constructed a multi-dimensional, orthogonal, and anti-hacking reward system.
The final composite reward signal $\mathcal{R}_{\text{total}}$ is a weighted fusion of feedback from three distinct dimensions:
\begin{equation}
  \mathcal{R}_{\text{total}} = \lambda_{\text{STT}} \cdot \mathcal{R}_{\text{STT}} + \lambda_{\text{Pref}} \cdot \mathcal{R}_{\text{Pref}} + \lambda_{\text{SIM}} \cdot \mathcal{R}_{\text{SIM}}.
\end{equation}

\begin{figure}[t]
  \centering
  \hspace*{-3em}\includegraphics[width=0.8\linewidth]{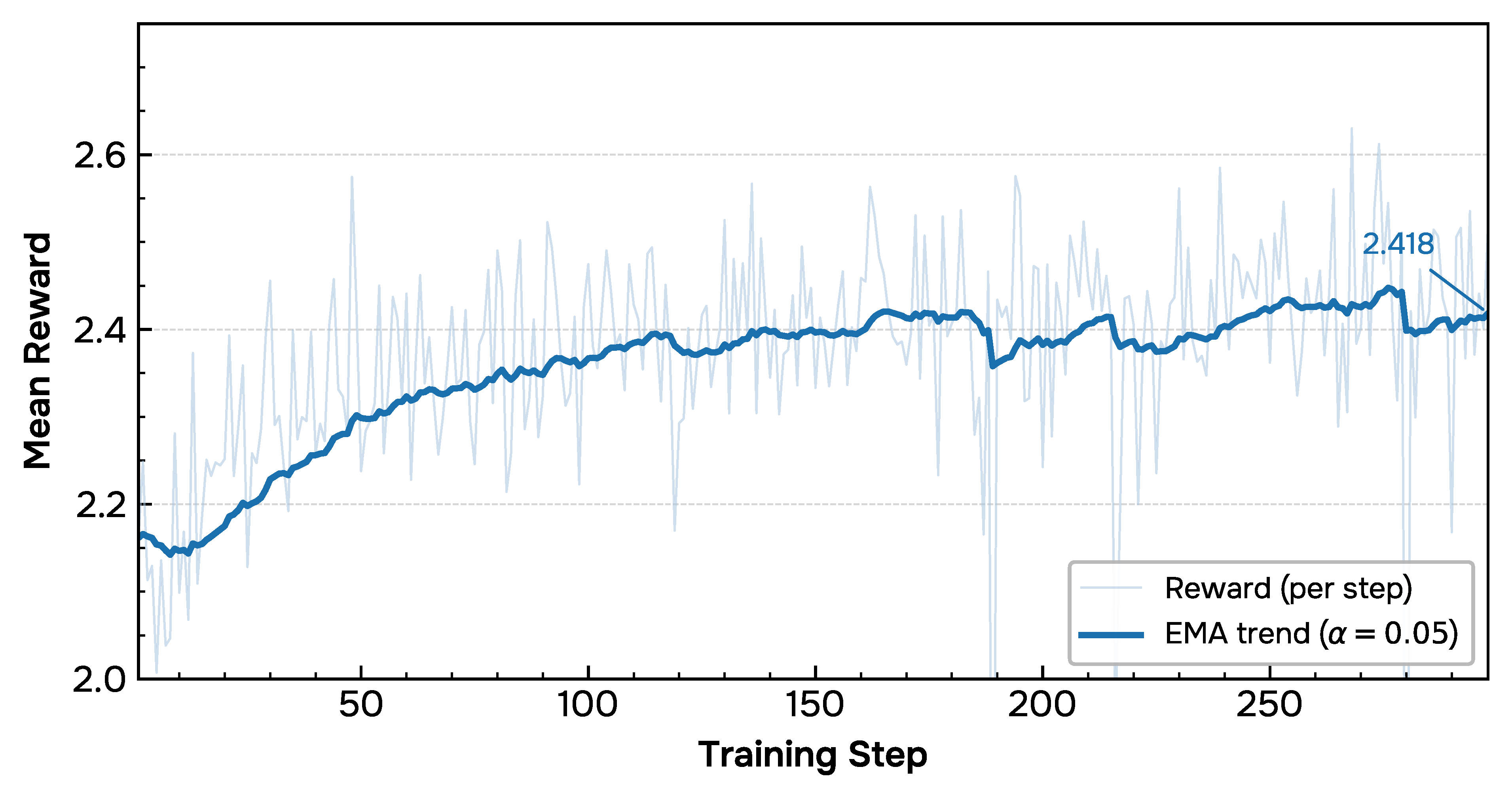}
  \caption{\textbf{Training reward curves during RL-based post-training.}}
  \label{fig:rl_loss}
\end{figure}

The semantic accuracy reward $\mathcal{R}_{\mathrm{STT}}$ utilizes the ASR caption model from the data pipeline (Section~\ref{sec:data}) that extracts per-token confidences as continuous signals.
To enforce strict instruction following, we implemented a token-weighted mask that applies substantially stronger penalties to incorrect speaker ID tags and additional penalties to missed vocal instructions.
The acoustic preference reward $\mathcal{R}_{\mathrm{Pref}}$ is scored by the speech quality model from the data pipeline (Section~\ref{sec:data}).
The timbre similarity reward $\mathcal{R}_{\mathrm{SIM}}$ utilizes an external voiceprint model to extract features and compute cosine similarity.
Figure~\ref{fig:rl_loss} shows that the total reward $\mathcal{R}_\text{total}$ (per-step mean over the batch) rises consistently before convergence, demonstrating the effectiveness of our multi-dimensional reward design in providing stable and coherent training signal throughout RL post-training.

In terms of system and optimization dynamics, to prevent computationally heavy scoring models from idling the primary node, we abstracted the entire scoring system into an asynchronous, decoupled architecture.
Combined with a centralized waveform cache, this maximizes the rollout throughput during the RL-based post-training phase.
To efficiently compute the KL divergence penalty in the policy loss without perpetually hosting a redundant full reference model in VRAM, we design a LoRA weight-swap mechanism: the reference policy is maintained as a LoRA weight backup in CPU memory and dynamically swapped in for gradient-free forward passes during divergence computation, significantly reducing peak memory footprint.
We employ rank-stabilized LoRA (rsLoRA, $r=16, \alpha=64$)~\citep{kalajdzievski2023rank}, updating exclusively the MLP layers.

\section{Inference Engine}\label{sec:inference}

To achieve high-throughput and ultra-low latency in production deployments, our inference engine is built upon SGLang~\citep{zheng2024sglangefficientexecutionstructured}, a state-of-the-art (SOTA) serving framework originally designed for LLMs.
SGLang provides a highly optimized execution backend featuring continuous batching, paged KV cache, CUDA graph replay, and notably, RadixAttention for efficient prefix caching.
By leveraging this advanced serving infrastructure, we can maximize GPU utilization and minimize generation latency. 

Notably, achieving this high performance does not require massive modifications to the underlying engine. 
As our Dual-AR architecture is structurally isomorphic to standard autoregressive text LLMs, the autoregressive complexity is fully encapsulated within the native forward pass.
The SGLang core scheduler and execution engine remain completely agnostic to the audio modality, allowing us to inherit all LLM-native optimizations with zero friction.
To adapt this framework for audio generation, we introduced serveral targeted modifications.

First, we implemented an I/O bypass at the API level (skipping the standard text tokenizer and detokenizer), allowing mixed prompts consisting of semantic inputs and discrete acoustic tokens, with streaming acoustic token ID outputs.
Second,We extend the original RadixCache, which was designed for indexing individual text tokens, by introducing minimal multi-token indexing keys that jointly encode semantic and acoustic tokens. This modification enables RadixCache to cache diverse reference audio contexts, significantly improving KV cache hit rates in real-world serving environments.Third, to maximize GPU utilization, we analyze the system bottlenecks and find that LLM decoding is predominantly memory-bandwidth bound. We therefore leverage MPS to co-schedule vocoder decoding with LLM decoding on the same GPU, enabling concurrent execution that improves system throughput while preserving low latency.

We evaluated the production-ready inference performance of Fish~Audio~S2 on a single NVIDIA H200 GPU. Thanks to the lightweight Dual-AR architecture and SGLang optimizations, the system achieves industry-leading generation metrics:

\begin{itemize}[leftmargin=2em, itemsep=-0.3em, topsep=-0.2em]
    \item \textbf{Real-Time Factor (RTF).} The model achieves an exceptional RTF of 0.195, generating high-fidelity audio vastly faster than real-time.
    \item \textbf{Time-to-First-Audio (TTFA).} Benefiting from audio tokenizer decoding co-scheduling and RadixCache hits, the system achieves a time-to-first-audio (TTFA) as low as 100 ms in the production serving environment.
    \item \textbf{Throughput.} Under high concurrency, the engine sustains a maximum throughput of 3000+ acoustic tokens per second while keeping the RTF below 0.5, enabling real-time streaming synthesis even under heavy load and demonstrating its readiness for large-scale production serving.
\end{itemize}

Beyond raw throughput and latency, this architectural alignment also enables highly efficient voice reuse in production.
As described in Section \ref{sec:training}, Fish~Audio~S2 inserts the deterministic reference-audio tokens into the system prompt.
During inference, SGLang’s Radix tree caches the corresponding KV states.
When the same voice is reused across multiple requests, this design delivers a high prefix-cache hit rate (86.4\% on average and over 90\% at peak).
As a result, repeated requests can largely skip the reference-audio prefill stage, making prompt-processing overhead nearly negligible.
\section{Evaluation}\label{sec:evaluation}

We conduct a comprehensive evaluation of Fish Audio S2 across two primary dimensions.
First, we utilize traditional objective metrics, focusing on WER, CER, and speaker similarity (SIM) to measure transcription accuracy and acoustic fidelity. Second, we employ an LLM-as-a-Judge benchmark to assess nuanced subjective qualities, specifically evaluating the model's instruction-following capabilities and overall speech naturalness.

\subsection{Objective Evaluation}

Our objective evaluation covers three components: (1) voice-cloning intelligibility on Seed-TTS-Eval (English and Chinese); (2) multilingual intelligibility and speaker similarity on CV3-Eval and the Minimax Multilingual Testset; and (3) long-form generation quality on a modified Long-TTS-Eval~\citep{wang2025mgm}.

\subsubsection{Seed-TTS-Eval}

\vspace{1em}

\begin{table}[!htbp]
  \centering
  
  \caption{\textbf{Results on Seed-TTS-Eval.}}
  \vspace{-2pt}
  \label{table:seed_eval}

  \setlength{\tabcolsep}{8pt}
  \renewcommand{\arraystretch}{1.12}
  \resizebox{0.55\textwidth}{!}{
  \begin{tabular}{lc}
    \toprule
    \textbf{Model} & \begin{tabular}[c]{@{}c@{}}\textbf{WER} (\%) $\downarrow$ \\ \textit{test-zh \textbar{} test-en \textbar{} zh-hard}\end{tabular} \\
    \midrule
    \textbf{Fish Audio S2} & \makebox[2.3em][c]{\textbf{0.54}} \textbar{} \makebox[2.3em][c]{\textbf{0.99}} \textbar{} \makebox[2.8em][c]{5.99} \\
    Fish Audio S1 & \makebox[2.3em][c]{\textbf{0.54}} \textbar{} \makebox[2.3em][c]{1.07} \textbar{} \makebox[2.8em][c]{17.00} \\
    CosyVoice 3-1.5B (\citealp{du2025cosyvoice3inthewildspeech}) & \makebox[2.3em][c]{1.12} \textbar{} \makebox[2.3em][c]{2.21} \textbar{} \makebox[2.8em][c]{\textbf{5.83}} \\
    Qwen3-TTS (\citealp{hu2026qwen3ttstechnicalreport}) & \makebox[2.3em][c]{0.77} \textbar{} \makebox[2.3em][c]{1.24} \textbar{} \makebox[2.8em][c]{--} \\
    Qwen3-Omni (\citealp{xu2025qwen3omnitechnicalreport}) & \makebox[2.3em][c]{1.07} \textbar{} \makebox[2.3em][c]{1.39} \textbar{} \makebox[2.8em][c]{--} \\
    FireRedTTS-2 (\citealp{xie2025fireredtts2longconversationalspeech}) & \makebox[2.3em][c]{1.14} \textbar{} \makebox[2.3em][c]{1.95} \textbar{} \makebox[2.8em][c]{--} \\
    Seed-TTS (\citealp{anastassiou2024seedttsfamilyhighqualityversatile}) & \makebox[2.3em][c]{1.12} \textbar{} \makebox[2.3em][c]{2.25} \textbar{} \makebox[2.8em][c]{7.59} \\
    Minimax Speech-02 (\citealp{zhang2025minimaxspeechintrinsiczeroshottexttospeech}) & \makebox[2.3em][c]{0.99} \textbar{} \makebox[2.3em][c]{1.90} \textbar{} \makebox[2.8em][c]{--} \\
    \bottomrule
  \end{tabular}
  }
\end{table}

\vspace{1em}

We assess voice-cloning intelligibility on Seed-TTS-Eval using WER over the test-zh, test-en, and ZH-hard splits. WER is computed by transcribing synthesized audio with Whisper-large-v3~\citep{radford2023robust} for English and Paraformer-zh~\cite{gao2023funasr} for Chinese, following the benchmark protocol; results are reported in Table~\ref{table:seed_eval}.

Compared with other open-source and closed-source models, Fish Audio S2 achieves leading WER on both Chinese and English, while remaining competitive on ZH-hard.
This indicates clearer and more stable pronunciation in the voice-cloning task.

\subsubsection{Multilingual Evaluation}

To evaluate multilingual correctness, we test on the Minimax Multilingual Testset and CV3-Eval.
The CV3-Eval benchmark covers 9 major languages and is designed to comprehensively measure zero-shot speech synthesis and expressive voice cloning performance in unrestrained, in-the-wild settings.
The Minimax Multilingual Testset covers 24 major languages and is designed to comprehensively measure speech synthesis and zero-shot voice cloning performance in multilingual settings.

As shown in Table~\ref{minimax_test} and Table~\ref{cv3_test}, Fish Audio S2 demonstrates strong multilingual intelligibility, robust speaker preservation, and high content fidelity in zero-shot voice cloning.
On the Minimax Multilingual Testset, it achieves the lowest WER in 11 out of 24 languages and the highest SIM in 17 out of 24.
This trend is further validated on the 9-language CV3-Eval subset, where Fish Audio S2 attains the best error rate across all reported languages, outperforming CosyVoice variants by clear margins and reducing the average error from 3.96 to 3.01 (a 23.9\% relative reduction) compared to Fish Audio S1.
Performance gains are particularly consistent in Chinese, English, Japanese, Korean, and major European languages.
While MiniMax-Speech and ElevenLabs still maintain an advantage in certain low-resource languages (typically those with under 1,000 hours of training data), Fish Audio S2 remains competitive in intelligibility and frequently achieves superior SIM, highlighting its stronger cross-lingual timbre consistency.

\vspace{1em}

\begin{table}[H]
  \centering
  \caption{\textbf{Results on CV3-Eval multilingual voice cloning subset.}}
  \label{cv3_test}
  \resizebox{0.87\textwidth}{!}{
  \setlength{\tabcolsep}{6pt}
  \renewcommand{\arraystretch}{1.12}
  \begin{tabular}{lccccccccccc}
    \toprule
    \multirow{2}{*}{\textbf{Model}} & \multicolumn{11}{c}{\textbf{WER} (\%) $\downarrow$} \\[-1pt]
    \cmidrule{2-12}
    & zh & en & ja & ko & de & es & fr & it & ru & hard-zh & hard-en \\[-1pt]
    \midrule
    \textbf{Fish Audio S2} & \textbf{2.65} & \textbf{2.43} & \textbf{3.96} & \textbf{2.76} & \textbf{2.22} & \textbf{2.00} & \textbf{6.26} & \textbf{2.04} & \textbf{2.78} & 9.10 & \textbf{4.40} \\
    Fish Audio S1 & 2.98 & 3.00 & 4.54 & 3.19 & 2.78 & 2.84 & 8.15 & 2.94 & 5.18 & 9.72 & 7.26 \\
    CosyVoice2 + DiffRO & 3.00 & 4.72 & 6.36 & 5.14 & -- & -- & -- & -- & -- & 10.66 & 10.25 \\
    CosyVoice3-0.5B + DiffRO & 2.89 & 3.68 & 5.15 & 4.02 & 4.51 & 2.99 & 8.56 & 2.94 & 3.79 & \textbf{8.26} & 7.60 \\
    CosyVoice3-1.5B + DiffRO & 3.01 & 3.71 & 5.27 & 4.01 & 3.93 & 3.26 & 8.09 & 2.72 & 4.11 & 9.06 & 7.56 \\
    \bottomrule
  \end{tabular}
  }
\end{table}

\begin{table}[H]
  \centering
  \caption{\textbf{Results on the Minimax Multilingual Testset.}}
  \label{minimax_test}
  \resizebox{\textwidth}{!}{
  \setlength{\tabcolsep}{4pt}
  \renewcommand{\arraystretch}{1.12}
  \begin{tabular}{l cccc cccc}
    \toprule
    \multirow{2}{*}{\textbf{Language}} & \multicolumn{4}{c}{\textbf{WER} (\%) $\downarrow$} & \multicolumn{4}{c}{\textbf{SIM} $\uparrow$} \\
    \cmidrule(lr){2-5}\cmidrule(lr){6-9}
    & MiniMax & ElevenLabs & \textbf{Fish Audio S2} & Fish Audio S1 & MiniMax & ElevenLabs & \textbf{Fish Audio S2} & Fish Audio S1 \\
    \midrule
    Arabic      & \textbf{1.665} & 1.666 & 3.500 & 6.420 & 0.736 & 0.706 & \textbf{0.750} & 0.713 \\
    Cantonese   & 34.111 & 51.513 & \textbf{30.670} & 48.800 & 0.778 & 0.670 & \textbf{0.805} & 0.773 \\
    Chinese     & 2.252 & 16.026 & \textbf{0.730} & 0.980 & 0.780 & 0.677 & \textbf{0.816} & 0.777 \\
    Czech       & 3.875 & \textbf{2.108} & 2.840 & 18.020 & 0.796 & 0.685 & \textbf{0.798} & 0.733 \\
    Dutch       & 1.143 & \textbf{0.803} & 0.990 & 2.270 & \textbf{0.738} & 0.680 & 0.730 & 0.701 \\
    English     & 2.164 & 2.339 & \textbf{1.620} & 2.370 & 0.756 & 0.613 & \textbf{0.797} & 0.785 \\
    Finnish     & 4.666 & \textbf{2.964} & 3.330 & 10.000 & \textbf{0.835} & 0.759 & 0.819 & 0.780 \\
    French      & 4.099 & 5.216 & \textbf{3.050} & 5.700 & 0.628 & 0.535 & \textbf{0.698} & 0.602 \\
    German      & 1.906 & 0.572 & \textbf{0.550} & 0.650 & 0.733 & 0.614 & \textbf{0.767} & 0.719 \\
    Greek       & 2.016 & \textbf{0.991} & 5.740 & 19.840 & \textbf{0.826} & 0.733 & 0.795 & 0.746 \\
    Hindi       & 6.962 & \textbf{5.827} & 14.640 & 32.280 & 0.818 & 0.730 & \textbf{0.821} & 0.792 \\
    Indonesian  & 1.237 & \textbf{1.059} & 1.460 & 8.000 & 0.729 & 0.660 & \textbf{0.763} & 0.696 \\
    Italian     & 1.543 & 1.743 & \textbf{1.270} & 1.870 & 0.699 & 0.579 & \textbf{0.747} & 0.714 \\
    Japanese    & 3.519 & 10.646 & \textbf{2.760} & 3.450 & 0.776 & 0.738 & \textbf{0.796} & 0.780 \\
    Korean      & 1.747 & 1.865 & \textbf{1.180} & 1.990 & 0.776 & 0.700 & \textbf{0.817} & 0.747 \\
    Polish      & 1.415 & \textbf{0.766} & 1.260 & 3.410 & 0.802 & 0.729 & \textbf{0.819} & 0.794 \\
    Portuguese  & 1.877 & 1.331 & \textbf{1.140} & 1.940 & \textbf{0.805} & 0.711 & 0.781 & 0.756 \\
    Romanian    & 2.878 & \textbf{1.347} & 10.740 & 19.490 & \textbf{0.809} & 0.699 & 0.733 & 0.739 \\
    Russian     & 4.281 & 3.878 & \textbf{2.400} & 5.250 & 0.761 & 0.676 & \textbf{0.790} & 0.764 \\
    Spanish     & 1.029 & 1.084 & \textbf{0.910} & 1.780 & 0.762 & 0.615 & \textbf{0.776} & 0.753 \\
    Thai        & \textbf{2.701} & 73.936 & 4.230 & 96.750 & \textbf{0.800} & 0.588 & 0.786 & 0.691 \\
    Turkish     & 1.520 & \textbf{0.699} & 0.870 & 2.260 & 0.779 & 0.596 & \textbf{0.835} & 0.786 \\
    Ukrainian   & 1.082 & \textbf{0.997} & 2.300 & 14.490 & 0.730 & 0.647 & \textbf{0.747} & 0.653 \\
    Vietnamese  & \textbf{0.880} & 73.415 & 7.410 & 78.130 & \textbf{0.743} & 0.369 & 0.740 & 0.664 \\
    \bottomrule
  \end{tabular}
  }
\end{table}

\FloatBarrier 

\subsubsection{Long-Audio Benchmark}

To evaluate long-form speech generation, we adopt the Long-TTS-Eval 
dataset, which covers six content categories—literature, news, knowledge, speeches, reviews, and academic papers—in both English and Chinese, sourced from news outlets, Wikipedia, YouTube transcripts, and arXiv papers. Reference audios are sampled from the Seed-TTS-Eval benchmark.
Since our model is pretrained to support a maximum of 8,192 context length (Section~\ref{sec:pretrain}), samples exceeding this context limit are truncated at sentence boundaries.
We first estimate the ratio of generated audio tokens to input text tokens from a subset of Long-TTS-Eval, then set text token limits of 939 for English and 884 for Chinese, such that the expected generated audio does not exceed approximately 185 seconds.
To increase length diversity, we apply a $\pm$30\% random perturbation around these thresholds.
The final benchmark contains both truncated and non-truncated samples, with text token counts ranging from 74 to 1,211 (mean: 760) for English and 32 to 1,146 (mean: 589) for Chinese.

We follow the Seed-TTS-Eval protocol and report WER for English and CER for Chinese, computed using Whisper-large-v3 and Paraformer-zh with 28s none-overlapped chunks, transcribing each chunk separately, and then concatenate them to obtain the final transcription, respectively.
To assess speaker consistency over long utterances, we compute speaker similarity using WavLM-large~\citep{chen2022wavlm}. 
Specifically, the generated audio is segmented into 3-second chunks with a 1.5-second hop; speaker embeddings are extracted for each chunk and compared against the reference audio via cosine similarity.
We report the mean (SIM-Mean) and standard deviation (SIM-Std) across all chunks, where a low SIM-Std indicates stable timbre throughout the utterance.
As shown in Table~\ref{tab:long-audio-results}, Fish~Audio~S2 achieves the lowest WER and CER among all evaluated models on both English and Chinese, demonstrating its robustness in generating long audio.

\begin{table}[t]
\centering
\caption{\textbf{Results on the Long-Audio benchmark.}}
\label{tab:long-audio-results}
\resizebox{0.95\textwidth}{!}{
  \setlength{\tabcolsep}{8pt}
  \renewcommand{\arraystretch}{1.12}
\begin{tabular}{lcccccc}
\toprule
\multirow{2}{*}{\textbf{Model}} & \multicolumn{3}{c}{\textbf{English}} & \multicolumn{3}{c}{\textbf{Chinese}} \\
\cmidrule(lr){2-4} \cmidrule(lr){5-7}
& \textbf{WER} (\%) $\downarrow$ & \textbf{SIM-Mean} $\uparrow$ & \textbf{SIM-Std} $\downarrow$ 
& \textbf{CER} (\%) $\downarrow$ & \textbf{SIM-Mean} $\uparrow$ & \textbf{SIM-Std} $\downarrow$ \\
\midrule
\textbf{Fish Audio S2}       & \textbf{4.38} & \textbf{0.523} & 0.0761 & \textbf{5.95} & 0.557 & 0.0923 \\
Fish Audio S1       & 6.26 & 0.436 & 0.108 & 6.44 & 0.505 & 0.108 \\
Qwen3-TTS~(\citealp{hu2026qwen3ttstechnicalreport})         & 7.69 & 0.390 & 0.0737 & 8.09 & 0.574 & 0.0614 \\
VibeVoice~(\citealp{peng2025vibevoicetechnicalreport})         & 28.0 & 0.530 & \textbf{0.0572} & 26.2 & \textbf{0.609} & \textbf{0.0537} \\
\bottomrule
\end{tabular}
}
\end{table}

\subsection{LLM-as-a-Judge Evaluation}

While objective metrics provide a foundational assessment of model accuracy, they often fall short in capturing nuanced generation qualities such as naturalness, prosody, and semantic coherence.
To complement human evaluation at scale, we introduce comprehensive LLM-as-a-Judge evaluation.
Specifically, we systematically assess the model's generative capabilities across three distinct benchmarks: (1) human-level indistinguishability on the Audio Turing Test; (2) advanced synthesis behaviors on the Emergent TTS Eval; and (3) fine-grained instruction following and controllability on our Fish-Instruction-Benchmark.
This automated approach ensures a highly scalable and reproducible evaluation pipeline while providing insights that strongly correlate with human perception.

\subsubsection{Audio Turing Test}

To carefully evaluate whether our generated speech achieves human-level indistinguishability, we adopt the Audio Turing Test (ATT) framework.
Traditional Mean Opinion Score (MOS) evaluations often suffer from human rater subjectivity, anchoring effects, and a lack of cross-context comparability.
To overcome these limitations, the ATT framework simplifies the assessment into a Turing Test-inspired paradigm.
Instead of relying on complex and biased continuous scales, it requires the evaluator to make a straightforward ternary classification: \textit{Human}, \textit{Machine}, or \textit{Unclear}.
Within our LLM-as-a-Judge pipeline, we utilize the multi-dimensional ATT-Corpus to systematically assess the model's capability in synthesizing highly realistic and contextually appropriate speech, effectively mitigating the scaling biases inherent in traditional subjective metrics.

\begin{table}[b]
  \centering
  \caption{\textbf{Posterior summary statistics from the ATT evaluation: means, standard deviations (Std), and 95\% highest density intervals (HDI).}}
  \label{tab:att_posterior}
  \resizebox{0.63\textwidth}{!}{
  \setlength{\tabcolsep}{8pt}
  \renewcommand{\arraystretch}{1.12}
  \begin{tabular}{l l l}
    \toprule
    \textbf{Model} & \textbf{Mean (Std)} & \textbf{95}\% \textbf{HDI} \\
    \midrule
    \textbf{Fish Audio S2} & 0.483 (0.068) & [0.477, 0.489] \\
    \textbf{Fish Audio S2 (w/ instruction)} & \textbf{0.515} (0.061) & \textbf{[0.510, 0.521]}  \\
    Fish Audio S1 & 0.479 (0.087) & [0.471, 0.486] \\
    Seed-TTS~(\citealp{anastassiou2024seedttsfamilyhighqualityversatile}) & 0.417 (0.011) & [0.398, 0.438] \\
    MiniMax-Speech~(\citealp{zhang2025minimaxspeechintrinsiczeroshottexttospeech}) & 0.387 (0.011) & [0.368, 0.407] \\
    Step-Audio~(\citealp{huang2025stepaudiounifiedunderstandinggeneration}) & 0.286 (0.011) & [0.266, 0.307] \\
    CosyVoice~(\citealp{du2024cosyvoice}) & 0.234 (0.010) & [0.214, 0.254] \\
    GPT-4o & 0.138 (0.011) & [0.118, 0.158] \\
    \bottomrule
  \end{tabular}
  }
\end{table}

First, we utilize Gemini-3-Pro to perform instruction expansion on all 499 audio samples.
Then, we select 5 prompt audios from the Seed-TTS-Eval benchmark to synthesize speech for both the original texts and the rewritten instruction prompts.
The generated audio clips are then evaluated using the Auto-ATT model provided in the original ATT paper.
The final quantitative results are summarized in Table~\ref{tab:att_posterior}.

The evaluation indicates that Fish Audio S2 significantly outperforms the recently released models under both experimental settings.
Notably, in the instruction-rewritten setting, our model surpasses the previous SOTA by ~30\%, establishing a new industry benchmark.
Specifically, Fish Audio S2 eclipses previous SOTA performance even in the baseline synthesis setting (using original texts), demonstrating the model's foundational strength in generating highly natural and realistic speech.
Furthermore, the audio synthesized from the LLM-rewritten instructions yields substantial improvements over the unrewritten baselines. This observation highlights that instruction following plays a critical role in enhancing speech authenticity and naturalness, while rigorously validating the robust instruction-following capabilities of the Fish Audio S2 model.

\begin{table}[t]
\centering
\caption{\textbf{Results on EmergentTTS-Eval across five dimensions, with WER and Win-Rate against baseline.}$\star$ indicates strong prompting.}
\label{table:emergent_test}
\resizebox{\textwidth}{!}{
\renewcommand{\arraystretch}{1.12}

\begin{tabular}{lcccccccccccc}
\toprule
\textbf{Model} & 
\multicolumn{2}{c}{\textbf{Overall}} & 
\multicolumn{2}{c}{\textbf{Emotions}} & 
\multicolumn{2}{c}{\textbf{Foreign Words}} & 
\multicolumn{2}{c}{\textbf{Paralinguistics}} & 
\multicolumn{2}{c}{\textbf{Questions}} & 
\multicolumn{2}{c}{\textbf{Syntactic Comp.}} \\
\cmidrule(lr){2-3} \cmidrule(lr){4-5} \cmidrule(lr){6-7} \cmidrule(lr){8-9} \cmidrule(lr){10-11} \cmidrule(lr){12-13}
  & \textbf{WER} (\%) $\downarrow$ & \textbf{Win} (\%) $\uparrow$ & \textbf{WER} (\%) $\downarrow$ & \textbf{Win} (\%) $\uparrow$ & \textbf{WER} (\%) $\downarrow$ & \textbf{Win} (\%) $\uparrow$ & \textbf{WER} (\%) $\downarrow$ & \textbf{Win} (\%) $\uparrow$ & \textbf{WER} (\%) $\downarrow$ & \textbf{Win} (\%) $\uparrow$ & \textbf{WER} (\%) $\downarrow$ & \textbf{Win} (\%) $\uparrow$ \\
\midrule
\textbf{Fish Audio S2} $\star$ & 8.15 & \underline{\textbf{81.88}} & \underline{\textbf{0.52}} & 86.61 & 14.74 & 63.39 & 24.58 & \underline{\textbf{91.61}} & 0.38 & \underline{\textbf{84.41}} & 0.51 & \underline{\textbf{83.39}} \\
Fish Audio S1 & 7.60 & 36.88 & 0.75 & 24.82 & 15.77 & 53.75 & 20.02 & 37.68 & 0.72 & 43.55 & 0.74 & 24.64 \\
\midrule
Gemini-2.5-Flash-Preview-TTS $\star$ & \underline{\textbf{6.35}} & 79.10 & 0.71 & \underline{\textbf{97.32}} & \underline{\textbf{11.80}} & 65.56 & 18.38 & 91.25 & 0.40 & 74.82 & \underline{\textbf{0.50}} & 67.14 \\
Gemini-2.5-Flash-Preview-TTS & 6.59 & 78.00 & 0.60 & 95.00 & 12.99 & 63.03 & 18.25 & 89.10 & 0.36 & 74.82 & 0.73 & 68.03 \\
gpt-4o-audio-preview $\star$ & 7.75 & 77.36 & 1.82 & 96.96 & 13.30 & 70.00 & 21.15 & 89.46 & 1.38 & 62.50 & 1.16 & 68.39 \\
Gemini-2.5-Pro-Preview-TTS $\star$ & 7.91 & 73.00 & 0.87 & 92.14 & 16.22 & 63.44 & 20.87 & 84.28 & 0.72 & 65.71 & 0.87 & 60.00 \\
gpt-4o-mini-audio-preview $\star$ & 8.86 & 66.27 & 9.34 & 74.10 & 12.70 & \underline{\textbf{73.39}} & 20.92 & 68.03 & 0.74 & 55.53 & 0.72 & 60.35 \\
gpt-4o-audio-preview $\star$ & 7.64 & 64.90 & 0.93 & 67.14 & 13.75 & 71.60 & 20.56 & 77.32 & 1.72 & 49.64 & 1.26 & 58.92 \\
gpt-4o-mini-tts $\star$ & 7.11 & 61.98 & 0.71 & 67.32 & 12.07 & 63.03 & 21.33 & 63.39 & 0.66 & 54.28 & 0.84 & 61.96 \\
gpt-4o-audio-preview & 8.27 & 60.53 & 1.03 & 56.78 & 14.72 & 68.75 & 23.16 & 71.60 & 1.19 & 48.75 & 1.25 & 56.78 \\
gpt-4o-mini-audio-preview & 7.18 & 55.15 & 0.95 & 62.90 & 14.48 & 64.82 & 19.04 & 52.14 & 0.55 & 45.71 & 0.88 & 50.17 \\
\midrule
\underline{Baseline} gpt-4o-mini-tts & 7.23 & 50.00 & 0.72 & -- & 13.45 & -- & 20.55 & -- & 0.42 & -- & 1.04 & -- \\
\midrule
HumeAI $\star$ & 8.62 & 49.12 & 0.83 & 71.78 & 21.05 & 41.25 & 19.84 & 42.14 & 0.38 & 42.14 & 0.93 & 48.57 \\
Higgs Audio V2~(\citealp{higgsaudio2025}) & 11.82 & 44.64 & 0.97 & 69.28 & 22.26 & 21.42 & 31.34 & 45.53 & 3.23 & 45.35 & 1.29 & 41.60 \\
minimax/speech-02-hd~(\citealp{zhang2025minimaxspeechintrinsiczeroshottexttospeech}) & 6.79 & 40.96 & 0.57 & 41.60 & 14.58 & 31.96 & \underline{\textbf{17.69}} & 33.57 & \underline{\textbf{0.27}} & 52.14 & 0.84 & 45.53 \\
11Labs eleven multilingual v2 & 7.64 & 36.96 & 0.63 & 35.35 & 14.44 & 36.60 & 21.51 & 52.14 & 0.49 & 28.21 & 1.15 & 32.50 \\
Qwen 2.5 Omni~(\citealp{xu2025qwen25omnitechnicalreport}) $\star$ & 18.34 & 32.94 & 2.41 & 46.60 & 26.77 & 14.46 & 58.44 & 21.25 & 0.87 & 48.92 & 3.47 & 33.75 \\
Orpheus TTS & 13.63 & 32.56 & 1.81 & 39.06 & 22.31 & 14.64 & 40.94 & 48.57 & 1.48 & 31.07 & 1.63 & 29.46 \\
Qwen 2.5 Omni & 20.02 & 30.25 & 1.22 & 41.07 & 26.98 & 12.50 & 57.48 & 20.89 & 12.77 & 49.10 & 1.66 & 27.67 \\
ResembleAI Chatterbox~(\citealp{chatterboxtts2025}) & 8.20 & 27.39 & 1.18 & 28.03 & 17.59 & 24.64 & 20.64 & 17.32 & 0.65 & 51.07 & 0.96 & 15.89 \\
Kokoro-82M & 10.49 & 25.46 & 0.71 & 18.03 & 22.17 & 13.21 & 28.37 & 5.89 & 0.56 & 43.39 & 0.65 & 46.78 \\
DeepGram Aura-2 & 10.24 & 25.21 & 3.45 & 17.50 & 21.41 & 15.89 & 23.73 & 20.89 & 1.24 & 43.03 & 1.36 & 28.75 \\
KyutAI-TTS~(\citealp{zeghidour2025streaming}) & 9.10 & 24.28 & 0.83 & 31.78 & 16.47 & 13.39 & 26.41 & 24.28 & 0.61 & 35.17 & 1.19 & 16.78 \\
MiniCPM-o~(\citealp{yao2024minicpm}) & 22.52 & 19.75 & 12.36 & 22.85 & 33.46 & 5.89 & 58.48 & 21.60 & 5.21 & 22.14 & 3.08 & 26.25 \\
F5TTS~(\citealp{chen2025f5}) & 11.93 & 19.43 & 0.70 & 29.64 & 23.51 & 3.21 & 31.66 & 19.82 & 1.62 & 19.46 & 2.14 & 25.00 \\
\bottomrule
\end{tabular}
}\end{table}

\subsubsection{EmergentTTS-Eval}

EmergentTTS-Eval is a comprehensive benchmark specifically designed to assess the capabilities of TTS models in handling complex linguistic, prosodic, and expressive challenges.
Unlike standard evaluations, this benchmark employs a large audio language model to assess TTS performance across six fine-grained and challenging scenarios.

To demonstrate our model's strong fine-grained instruction-following capability, we first use Gemini 3 Pro to rewrite all benchmark texts and then synthesize speech from the rewritten prompts.
Following the official EmergentTTS-Eval guidelines, we use Gemini 2.5 Pro as an AI judge to run side-by-side comparisons, measuring our audio against a gpt-4o-mini-tts baseline.

As shown in Table~\ref{table:emergent_test}, Fish Audio S2 achieves the highest overall win rate at 81.88\%, outperforming all listed systems and exceeding the 50\% baseline margin by +31.88 points.
Although its overall WER (8.15) is not the lowest among all models, it consistently delivers stronger perceived quality in instruction-sensitive scenarios, with leading win rates in paralinguistics (91.61\%), questions (84.41\%), and syntactic complexity (83.39\%), while also remaining highly competitive on emotions (86.61\%).
These results indicate that our model better converts complex textual instructions into expressive and controllable speech, confirming that instruction-following alignment is a key driver of real-world TTS preference.

\subsection{Fish Audio Instruction Benchmark}

As detailed in Appendix~\ref{app:benchmark}, we introduce the Fish Instruction Benchmark to evaluate fine-grained instruction following beyond conventional WER/MOS metrics.
Instead of relying on global style prompts, this benchmark uses inline vocal tags at specific word positions in both English and Chinese samples.
We then assess generated speech with Gemini 3 Pro using three complementary dimensions: Tag Activation Rate, Acoustic Naturalness, and Global Expressiveness.
We evaluate both Fish Audio S2 and Fish Audio S1 on this benchmark, and summarize the results in Table~\ref{table:fish-instruction-benchmark}. 

\begin{table}[h]
  \centering
  
  \caption{\textbf{Results on the Fish Audio Instruction Benchmark.}}
  \label{table:fish-instruction-benchmark}
  \resizebox{0.58\textwidth}{!}{
  \setlength{\tabcolsep}{8pt}
  \renewcommand{\arraystretch}{1.12}
  \begin{tabular}{l l c c}
    \toprule
    \textbf{Dataset} & \textbf{Metric} & \textbf{Fish Audio S2} & \textbf{Fish Audio S1} \\
    \midrule
    \multirow{3}{*}{Chinese}
      & TAR $\uparrow$ & \textbf{0.984} & 0.942 \\
      & Naturalness $\uparrow$ & \textbf{4.40} & 4.15 \\
      & Expressiveness $\uparrow$ & \textbf{4.94} & 4.65 \\
    \midrule
    \multirow{3}{*}{English}
      & TAR $\uparrow$ & \textbf{0.881} & 0.626 \\
      & Naturalness $\uparrow$ & \textbf{4.21} & 3.71 \\
      & Expressiveness $\uparrow$ & \textbf{4.50} & 3.93 \\

    \bottomrule
  \end{tabular}
  }
\end{table}

As shown in Table~\ref{table:fish-instruction-benchmark}, Fish Audio S2 consistently outperforms Fish Audio S1 across all reported instruction-following metrics in both Chinese and English settings.
On the Chinese set, TAR, Naturalness, and Expressiveness improve from 0.942/4.15/4.65 to 0.984/4.40/4.94.
On the English set, the improvements are larger, increasing from 0.626/3.71/3.93 to 0.881/4.21/4.50.
These results suggest that Fish Audio S2 provides more reliable tag activation and more natural, expressive vocal tag rendering under zero-shot instruction following.

\section{Conclusion}\label{sec:conclusion}

In this report, we present Fish Audio S2, a SOTA TTS system that supports fine-grained natural language control, long-form coherent synthesis, and native multi-speaker multi-turn generation with ultra-low RTF and TTFA in production.
These capabilities are enabled by three core contributions: a Dual-AR architecture that decouples temporal semantic modeling from depth-wise acoustic generation; a dual-purpose data pipeline in which the speech quality model and rich-transcription ASR model serve as both pre-training filters and RL reward signals, eliminating distribution shift between stages; and RL-based post-training with multi-dimensional rewards that jointly optimizes semantic accuracy, acoustic quality, and speaker similarity.
Evaluations on both objective benchmarks and LLM-as-a-Judge benchmarks demonstrate the effectiveness of these design choices across intelligibility, speaker similarity, naturalness, and instruction-following dimensions.
We additionally introduce the Fish Audio Instruction Benchmark to evaluate fine-grained tag-following beyond conventional WER and MOS metrics.
To facilitate research and broaden access to high-quality TTS, we open-source the model weights, fine-tuning code, and the SGLang-based inference engine.
We hope Fish Audio S2 serves as a strong open foundation for the next generation of expressive, controllable speech synthesis.

\section{Author}

\textbf{Core Contributors}

The core contributors are listed in order of contribution:

Shijia Liao, Yuxuan Wang, Songting Liu, Yifan Cheng

\textbf{Contributors}

All other contributors are sorted by last name and shuffled using Python 3.12 with random seed 42:

Ruoyi Zhang, Tianyu Li, Shidong Li, Yisheng Zheng, Xingwei Liu, Qingzheng Wang, Zhizhuo Zhou, Jiahua Liu, Xin Chen, Dawei Han

For correspondence, please contact oss@fish.audio.

\clearpage
\bibliography{fish_citations}

\begin{thebibliography}{45}
\providecommand{\natexlab}[1]{#1}
\providecommand{\url}[1]{\texttt{#1}}
\expandafter\ifx\csname urlstyle\endcsname\relax
  \providecommand{\doi}[1]{doi: #1}\else
  \providecommand{\doi}{doi: \begingroup \urlstyle{rm}\Url}\fi

\bibitem[Agarwal et~al.(2025)Agarwal, Ahmad, Ai, Altman, Applebaum, Arbus,
  Arora, Bai, Baker, Bao, et~al.]{agarwal2025gpt}
Sandhini Agarwal, Lama Ahmad, Jason Ai, Sam Altman, Andy Applebaum, Edwin
  Arbus, Rahul~K Arora, Yu~Bai, Bowen Baker, Haiming Bao, et~al.
\newblock gpt-oss-120b \& gpt-oss-20b model card.
\newblock \emph{arXiv preprint arXiv:2508.10925}, 2025.

\bibitem[Anastassiou et~al.(2024)Anastassiou, Chen, Chen, Chen, Chen, Chen,
  Cong, Deng, Ding, Gao, Gong, Huang, Huang, Huang, Huo, Jia, Li, Li, Li, Li,
  Li, Li, Liu, Liu, Liu, Liu, Liu, Liu, Lu, Pan, Wang, Wang, Wang, Wei, Wu,
  Yao, Yang, Yi, Zhang, Zhang, Zhang, Zhang, Zhang, Zhao, Zhong, and
  Zhuang]{anastassiou2024seedttsfamilyhighqualityversatile}
Philip Anastassiou, Jiawei Chen, Jitong Chen, Yuanzhe Chen, Zhuo Chen, Ziyi
  Chen, Jian Cong, Lelai Deng, Chuang Ding, Lu~Gao, Mingqing Gong, Peisong
  Huang, Qingqing Huang, Zhiying Huang, Yuanyuan Huo, Dongya Jia, Chumin Li,
  Feiya Li, Hui Li, Jiaxin Li, Xiaoyang Li, Xingxing Li, Lin Liu, Shouda Liu,
  Sichao Liu, Xudong Liu, Yuchen Liu, Zhengxi Liu, Lu~Lu, Junjie Pan, Xin Wang,
  Yuping Wang, Yuxuan Wang, Zhen Wei, Jian Wu, Chao Yao, Yifeng Yang, Yuanhao
  Yi, Junteng Zhang, Qidi Zhang, Shuo Zhang, Wenjie Zhang, Yang Zhang, Zilin
  Zhao, Dejian Zhong, and Xiaobin Zhuang.
\newblock Seed-tts: A family of high-quality versatile speech generation
  models, 2024.
\newblock URL \url{https://arxiv.org/abs/2406.02430}.

\bibitem[Barrault et~al.(2023)Barrault, Chung, Meglioli, Dale, Dong,
  Duppenthaler, Duquenne, Ellis, Elsahar, Haaheim,
  et~al.]{barrault2023seamless}
Lo{\"i}c Barrault, Yu-An Chung, Mariano~Coria Meglioli, David Dale, Ning Dong,
  Mark Duppenthaler, Paul-Ambroise Duquenne, Brian Ellis, Hady Elsahar, Justin
  Haaheim, et~al.
\newblock Seamless: Multilingual expressive and streaming speech translation.
\newblock \emph{arXiv preprint arXiv:2312.05187}, 2023.

\bibitem[{Boson AI}(2025)]{higgsaudio2025}
{Boson AI}.
\newblock {Higgs Audio V2: Redefining Expressiveness in Audio Generation}.
\newblock \url{https://github.com/boson-ai/higgs-audio}, 2025.
\newblock GitHub repository. Release blog available at
  \url{https://www.boson.ai/blog/higgs-audio-v2}.

\bibitem[Chen et~al.(2022)Chen, Wang, Chen, Wu, Liu, Chen, Li, Kanda, Yoshioka,
  Xiao, et~al.]{chen2022wavlm}
Sanyuan Chen, Chengyi Wang, Zhengyang Chen, Yu~Wu, Shujie Liu, Zhuo Chen, Jinyu
  Li, Naoyuki Kanda, Takuya Yoshioka, Xiong Xiao, et~al.
\newblock Wavlm: Large-scale self-supervised pre-training for full stack speech
  processing.
\newblock \emph{IEEE Journal of Selected Topics in Signal Processing},
  16\penalty0 (6):\penalty0 1505--1518, 2022.

\bibitem[Chen et~al.(2025)Chen, Niu, Ma, Deng, Wang, JianZhao, Yu, and
  Chen]{chen2025f5}
Yushen Chen, Zhikang Niu, Ziyang Ma, Keqi Deng, Chunhui Wang, JianZhao
  JianZhao, Kai Yu, and Xie Chen.
\newblock F5-tts: A fairytaler that fakes fluent and faithful speech with flow
  matching.
\newblock In \emph{Proceedings of the 63rd Annual Meeting of the Association
  for Computational Linguistics (Volume 1: Long Papers)}, pp.\  6255--6271,
  2025.

\bibitem[Cheng et~al.(2025)Cheng, Zhang, and Shi]{Cheng_2025}
Yifan Cheng, Ruoyi Zhang, and Jiatong Shi.
\newblock Miku-pal: An automated and standardized multimodal method for speech
  paralinguistic and affect labeling.
\newblock In \emph{Interspeech 2025}, pp.\  4308--4312. ISCA, August 2025.
\newblock \doi{10.21437/Interspeech.2025-648}.
\newblock URL \url{https://doi.org/10.21437/Interspeech.2025-648}.

\bibitem[D{\'e}fossez et~al.(2024)D{\'e}fossez, Mazar{\'e}, Orsini, Royer,
  P{\'e}rez, J{\'e}gou, Grave, and Zeghidour]{defossez2024moshi}
Alexandre D{\'e}fossez, Laurent Mazar{\'e}, Manu Orsini, Am{\'e}lie Royer,
  Patrick P{\'e}rez, Herv{\'e} J{\'e}gou, Edouard Grave, and Neil Zeghidour.
\newblock Moshi: a speech-text foundation model for real-time dialogue.
\newblock \emph{arXiv preprint arXiv:2410.00037}, 2024.

\bibitem[Du et~al.(2024)Du, Chen, Zhang, Hu, Lu, Yang, Hu, Zheng, Gu, Ma,
  et~al.]{du2024cosyvoice}
Zhihao Du, Qian Chen, Shiliang Zhang, Kai Hu, Heng Lu, Yexin Yang, Hangrui Hu,
  Siqi Zheng, Yue Gu, Ziyang Ma, et~al.
\newblock Cosyvoice: A scalable multilingual zero-shot text-to-speech
  synthesizer based on supervised semantic tokens.
\newblock \emph{arXiv preprint arXiv:2407.05407}, 2024.

\bibitem[Du et~al.(2025)Du, Gao, Wang, Yu, Zhao, Wang, Lv, Wang, Ni, Shi, An,
  Yang, Li, Chen, Gao, Chen, Gu, Chen, Chen, Zhang, Wang, and
  Ye]{du2025cosyvoice3inthewildspeech}
Zhihao Du, Changfeng Gao, Yuxuan Wang, Fan Yu, Tianyu Zhao, Hao Wang, Xiang Lv,
  Hui Wang, Chongjia Ni, Xian Shi, Keyu An, Guanrou Yang, Yabin Li, Yanni Chen,
  Zhifu Gao, Qian Chen, Yue Gu, Mengzhe Chen, Yafeng Chen, Shiliang Zhang, Wen
  Wang, and Jieping Ye.
\newblock Cosyvoice 3: Towards in-the-wild speech generation via scaling-up and
  post-training, 2025.
\newblock URL \url{https://arxiv.org/abs/2505.17589}.

\bibitem[Défossez et~al.(2022)Défossez, Copet, Synnaeve, and
  Adi]{défossez2022highfidelityneuralaudio}
Alexandre Défossez, Jade Copet, Gabriel Synnaeve, and Yossi Adi.
\newblock High fidelity neural audio compression, 2022.
\newblock URL \url{https://arxiv.org/abs/2210.13438}.

\bibitem[Gao et~al.(2023)Gao, Li, Wang, Luo, Shi, Chen, Li, Zuo, Du, Xiao,
  et~al.]{gao2023funasr}
Zhifu Gao, Zerui Li, Jiaming Wang, Haoneng Luo, Xian Shi, Mengzhe Chen, Yabin
  Li, Lingyun Zuo, Zhihao Du, Zhangyu Xiao, et~al.
\newblock Funasr: A fundamental end-to-end speech recognition toolkit.
\newblock \emph{arXiv preprint arXiv:2305.11013}, 2023.

\bibitem[Guo et~al.(2025)Guo, Yang, Zhang, Song, Wang, Zhu, Xu, Zhang, Ma, Bi,
  et~al.]{guo2025deepseek}
Daya Guo, Dejian Yang, Haowei Zhang, Junxiao Song, Peiyi Wang, Qihao Zhu,
  Runxin Xu, Ruoyu Zhang, Shirong Ma, Xiao Bi, et~al.
\newblock Deepseek-r1: Incentivizing reasoning capability in llms via
  reinforcement learning.
\newblock \emph{arXiv preprint arXiv:2501.12948}, 2025.

\bibitem[Hu et~al.(2026)Hu, Zhu, He, Guo, Zhang, Wang, Guo, Jiang, Hao, Guo,
  Zhang, Zhang, Yang, Xu, Zhou, and Lin]{hu2026qwen3ttstechnicalreport}
Hangrui Hu, Xinfa Zhu, Ting He, Dake Guo, Bin Zhang, Xiong Wang, Zhifang Guo,
  Ziyue Jiang, Hongkun Hao, Zishan Guo, Xinyu Zhang, Pei Zhang, Baosong Yang,
  Jin Xu, Jingren Zhou, and Junyang Lin.
\newblock Qwen3-tts technical report, 2026.
\newblock URL \url{https://arxiv.org/abs/2601.15621}.

\bibitem[Hu et~al.(2024)Hu, Tu, Han, He, Cui, Long, Zheng, Fang, Huang, Zhao,
  et~al.]{hu2024minicpm}
Shengding Hu, Yuge Tu, Xu~Han, Chaoqun He, Ganqu Cui, Xiang Long, Zhi Zheng,
  Yewei Fang, Yuxiang Huang, Weilin Zhao, et~al.
\newblock Minicpm: Unveiling the potential of small language models with
  scalable training strategies.
\newblock \emph{arXiv preprint arXiv:2404.06395}, 2024.

\bibitem[Huang et~al.(2025)Huang, Wu, Wang, Yan, Hu, Feng, Tian, Shen, Li,
  Chen, Liu, Miao, You, Chen, Yang, Huang, Zhang, Gong, Zhang, Zhou, Sun, Li,
  Feng, Wan, Hu, Wu, Zhen, Ming, Yuan, Zhang, Zhou, Li, Ma, Wang, An, Ji, Li,
  Wen, Kong, Ma, Liang, Mou, Ahmidi, Wang, Li, Miao, Xu, Wang, Shi, Sun, Hu,
  Sai, Liu, Huang, Yan, Wang, Jia, Zhang, Gong, Guo, Liu, Liu, Feng, Wu, Wu,
  Yang, Wang, Zhang, Lin, Li, Xia, Zhou, Zhao, Gu, Chen, Wu, Li, Li, Li, Liang,
  Wang, Hao, Wu, Tan, Sun, Shuai, Pang, Yang, Gao, Yuan, Liu, Deng, Jiang, Liu,
  Cao, Wang, Deng, Xie, Ming, He, Sun, Han, Huang, Deng, Liu, Wu, Zhao, Wei,
  Yu, Cao, Li, Ma, Xu, Wang, Shi, Wang, Zhou, Zhong, Zhang, Wei, Luo, Lu, Yin,
  Luo, Ding, Yan, Dai, Yang, Xie, Ge, Sun, Huang, Chang, Guan, Yang, Zhang,
  Jiao, Jiang, Shum, Chen, Li, Zhou, Zhang, Zhang, and
  Zhu]{huang2025stepaudiounifiedunderstandinggeneration}
Ailin Huang, Boyong Wu, Bruce Wang, Chao Yan, Chen Hu, Chengli Feng, Fei Tian,
  Feiyu Shen, Jingbei Li, Mingrui Chen, Peng Liu, Ruihang Miao, Wang You,
  Xi~Chen, Xuerui Yang, Yechang Huang, Yuxiang Zhang, Zheng Gong, Zixin Zhang,
  Hongyu Zhou, Jianjian Sun, Brian Li, Chengting Feng, Changyi Wan, Hanpeng Hu,
  Jianchang Wu, Jiangjie Zhen, Ranchen Ming, Song Yuan, Xuelin Zhang, Yu~Zhou,
  Bingxin Li, Buyun Ma, Hongyuan Wang, Kang An, Wei Ji, Wen Li, Xuan Wen,
  Xiangwen Kong, Yuankai Ma, Yuanwei Liang, Yun Mou, Bahtiyar Ahmidi, Bin Wang,
  Bo~Li, Changxin Miao, Chen Xu, Chenrun Wang, Dapeng Shi, Deshan Sun, Dingyuan
  Hu, Dula Sai, Enle Liu, Guanzhe Huang, Gulin Yan, Heng Wang, Haonan Jia,
  Haoyang Zhang, Jiahao Gong, Junjing Guo, Jiashuai Liu, Jiahong Liu, Jie Feng,
  Jie Wu, Jiaoren Wu, Jie Yang, Jinguo Wang, Jingyang Zhang, Junzhe Lin,
  Kaixiang Li, Lei Xia, Li~Zhou, Liang Zhao, Longlong Gu, Mei Chen, Menglin Wu,
  Ming Li, Mingxiao Li, Mingliang Li, Mingyao Liang, Na~Wang, Nie Hao, Qiling
  Wu, Qinyuan Tan, Ran Sun, Shuai Shuai, Shaoliang Pang, Shiliang Yang, Shuli
  Gao, Shanshan Yuan, Siqi Liu, Shihong Deng, Shilei Jiang, Sitong Liu,
  Tiancheng Cao, Tianyu Wang, Wenjin Deng, Wuxun Xie, Weipeng Ming, Wenqing He,
  Wen Sun, Xin Han, Xin Huang, Xiaomin Deng, Xiaojia Liu, Xin Wu, Xu~Zhao,
  Yanan Wei, Yanbo Yu, Yang Cao, Yangguang Li, Yangzhen Ma, Yanming Xu, Yaoyu
  Wang, Yaqiang Shi, Yilei Wang, Yizhuang Zhou, Yinmin Zhong, Yang Zhang,
  Yaoben Wei, Yu~Luo, Yuanwei Lu, Yuhe Yin, Yuchu Luo, Yuanhao Ding, Yuting
  Yan, Yaqi Dai, Yuxiang Yang, Zhe Xie, Zheng Ge, Zheng Sun, Zhewei Huang,
  Zhichao Chang, Zhisheng Guan, Zidong Yang, Zili Zhang, Binxing Jiao, Daxin
  Jiang, Heung-Yeung Shum, Jiansheng Chen, Jing Li, Shuchang Zhou, Xiangyu
  Zhang, Xinhao Zhang, and Yibo Zhu.
\newblock Step-audio: Unified understanding and generation in intelligent
  speech interaction, 2025.
\newblock URL \url{https://arxiv.org/abs/2502.11946}.

\bibitem[Kalajdzievski(2023)]{kalajdzievski2023rank}
Damjan Kalajdzievski.
\newblock A rank stabilization scaling factor for fine-tuning with lora.
\newblock \emph{arXiv preprint arXiv:2312.03732}, 2023.

\bibitem[Kong et~al.(2020)Kong, Kim, and Bae]{kong2020hifi}
Jungil Kong, Jaehyeon Kim, and Jaekyoung Bae.
\newblock Hifi-gan: Generative adversarial networks for efficient and high
  fidelity speech synthesis.
\newblock \emph{Advances in neural information processing systems},
  33:\penalty0 17022--17033, 2020.

\bibitem[Kumar et~al.(2023)Kumar, Seetharaman, Luebs, Kumar, and
  Kumar]{kumar2023high}
Rithesh Kumar, Prem Seetharaman, Alejandro Luebs, Ishaan Kumar, and Kundan
  Kumar.
\newblock High-fidelity audio compression with improved rvqgan.
\newblock \emph{Advances in Neural Information Processing Systems},
  36:\penalty0 27980--27993, 2023.

\bibitem[Li et~al.(2026)Li, Zhou, Wang, Wang, Wu, Zhou, Zhou, and
  Shu]{li2026indextts25technicalreport}
Yunpei Li, Xun Zhou, Jinchao Wang, Lu~Wang, Yong Wu, Siyi Zhou, Yiquan Zhou,
  and Jingchen Shu.
\newblock Indextts 2.5 technical report, 2026.
\newblock URL \url{https://arxiv.org/abs/2601.03888}.

\bibitem[Liao et~al.(2024{\natexlab{a}})Liao, Lan, and Zachariah]{liao2024eva}
Shijia Liao, Shiyi Lan, and Arun~George Zachariah.
\newblock Eva-gan: Enhanced various audio generation via scalable generative
  adversarial networks.
\newblock \emph{arXiv preprint arXiv:2402.00892}, 2024{\natexlab{a}}.

\bibitem[Liao et~al.(2024{\natexlab{b}})Liao, Wang, Li, Cheng, Zhang, Zhou, and
  Xing]{liao2024fishspeechleveraginglargelanguage}
Shijia Liao, Yuxuan Wang, Tianyu Li, Yifan Cheng, Ruoyi Zhang, Rongzhi Zhou,
  and Yijin Xing.
\newblock Fish-speech: Leveraging large language models for advanced
  multilingual text-to-speech synthesis, 2024{\natexlab{b}}.
\newblock URL \url{https://arxiv.org/abs/2411.01156}.

\bibitem[Lin et~al.(2017)Lin, Goyal, Girshick, He, and
  Doll{\'a}r]{lin2017focal}
Tsung-Yi Lin, Priya Goyal, Ross Girshick, Kaiming He, and Piotr Doll{\'a}r.
\newblock Focal loss for dense object detection.
\newblock In \emph{Proceedings of the IEEE international conference on computer
  vision}, pp.\  2980--2988, 2017.

\bibitem[Liu et~al.(2025)Liu, Chen, Li, Qi, Pang, Du, Lee, and
  Lin]{liu2025understanding}
Zichen Liu, Changyu Chen, Wenjun Li, Penghui Qi, Tianyu Pang, Chao Du, Wee~Sun
  Lee, and Min Lin.
\newblock Understanding r1-zero-like training: A critical perspective.
\newblock \emph{arXiv preprint arXiv:2503.20783}, 2025.

\bibitem[Manku et~al.(2025)Manku, Tang, Shi, Li, and
  Smola]{manku2025emergenttts}
Ruskin~Raj Manku, Yuzhi Tang, Xingjian Shi, Mu~Li, and Alex Smola.
\newblock Emergenttts-eval: Evaluating tts models on complex prosodic,
  expressiveness, and linguistic challenges using model-as-a-judge.
\newblock \emph{arXiv preprint arXiv:2505.23009}, 2025.

\bibitem[Peng et~al.(2025)Peng, Yu, Wang, Chang, Sun, Dong, Zhu, Xu, Bao, Wang,
  Huang, Xia, and Wei]{peng2025vibevoicetechnicalreport}
Zhiliang Peng, Jianwei Yu, Wenhui Wang, Yaoyao Chang, Yutao Sun, Li~Dong,
  Yi~Zhu, Weijiang Xu, Hangbo Bao, Zehua Wang, Shaohan Huang, Yan Xia, and Furu
  Wei.
\newblock Vibevoice technical report, 2025.
\newblock URL \url{https://arxiv.org/abs/2508.19205}.

\bibitem[Poria et~al.(2019)Poria, Hazarika, Majumder, Naik, Cambria, and
  Mihalcea]{poria2019meld}
Soujanya Poria, Devamanyu Hazarika, Navonil Majumder, Gautam Naik, Erik
  Cambria, and Rada Mihalcea.
\newblock Meld: A multimodal multi-party dataset for emotion recognition in
  conversations.
\newblock In \emph{Proceedings of the 57th annual meeting of the association
  for computational linguistics}, pp.\  527--536, 2019.

\bibitem[Radford et~al.(2023)Radford, Kim, Xu, Brockman, McLeavey, and
  Sutskever]{radford2023robust}
Alec Radford, Jong~Wook Kim, Tao Xu, Greg Brockman, Christine McLeavey, and
  Ilya Sutskever.
\newblock Robust speech recognition via large-scale weak supervision.
\newblock In \emph{International Conference on Machine Learning}, pp.\
  28492--28518. PMLR, 2023.

\bibitem[Rafailov et~al.(2023)Rafailov, Sharma, Mitchell, Manning, Ermon, and
  Finn]{rafailov2023direct}
Rafael Rafailov, Archit Sharma, Eric Mitchell, Christopher~D Manning, Stefano
  Ermon, and Chelsea Finn.
\newblock Direct preference optimization: Your language model is secretly a
  reward model.
\newblock \emph{Advances in neural information processing systems},
  36:\penalty0 53728--53741, 2023.

\bibitem[{Resemble AI}(2025)]{chatterboxtts2025}
{Resemble AI}.
\newblock {Chatterbox-TTS}.
\newblock \url{https://github.com/resemble-ai/chatterbox}, 2025.
\newblock GitHub repository.

\bibitem[Schulman et~al.(2017)Schulman, Wolski, Dhariwal, Radford, and
  Klimov]{schulman2017proximal}
John Schulman, Filip Wolski, Prafulla Dhariwal, Alec Radford, and Oleg Klimov.
\newblock Proximal policy optimization algorithms.
\newblock \emph{arXiv preprint arXiv:1707.06347}, 2017.

\bibitem[Shao et~al.(2024)Shao, Wang, Zhu, Xu, Song, Bi, Zhang, Zhang, Li, Wu,
  et~al.]{shao2024deepseekmath}
Zhihong Shao, Peiyi Wang, Qihao Zhu, Runxin Xu, Junxiao Song, Xiao Bi, Haowei
  Zhang, Mingchuan Zhang, YK~Li, Yang Wu, et~al.
\newblock {DeepSeekMath}: Pushing the limits of mathematical reasoning in open
  language models.
\newblock \emph{arXiv preprint arXiv:2402.03300}, 2024.

\bibitem[Shi et~al.(2025)Shi, Shim, and Watanabe]{shi2025uni}
Jiatong Shi, Hye-Jin Shim, and Shinji Watanabe.
\newblock Uni-versa: Versatile speech assessment with a unified network.
\newblock \emph{arXiv preprint arXiv:2505.20741}, 2025.

\bibitem[Wang et~al.(2025{\natexlab{a}})Wang, Zhong, Peng, Yang, Liu, Gui, Xia,
  Li, Yu, and Jia]{wang2025mgm}
Chengyao Wang, Zhisheng Zhong, Bohao Peng, Senqiao Yang, Yuqi Liu, Haokun Gui,
  Bin Xia, Jingyao Li, Bei Yu, and Jiaya Jia.
\newblock Mgm-omni: Scaling omni llms to personalized long-horizon speech.
\newblock \emph{arXiv preprint arXiv:2509.25131}, 2025{\natexlab{a}}.

\bibitem[Wang et~al.(2023)Wang, Chen, Wu, Zhang, Zhou, Liu, Chen, Liu, Wang,
  Li, He, Zhao, and Wei]{wang2023neuralcodeclanguagemodels}
Chengyi Wang, Sanyuan Chen, Yu~Wu, Ziqiang Zhang, Long Zhou, Shujie Liu, Zhuo
  Chen, Yanqing Liu, Huaming Wang, Jinyu Li, Lei He, Sheng Zhao, and Furu Wei.
\newblock Neural codec language models are zero-shot text to speech
  synthesizers, 2023.
\newblock URL \url{https://arxiv.org/abs/2301.02111}.

\bibitem[Wang et~al.(2025{\natexlab{b}})Wang, Zhao, Ren, Zhang, Li, Li, Wang,
  Qiu, Wan, Cao, et~al.]{wang2025att}
Xihuai Wang, Ziyi Zhao, Siyu Ren, Shao Zhang, Song Li, Xiaoyu Li, Ziwen Wang,
  Lin Qiu, Guanglu Wan, Xuezhi Cao, et~al.
\newblock Audio turing test: benchmarking the human-likeness of large language
  model-based text-to-speech systems in chinese.
\newblock \emph{arXiv preprint arXiv:2505.11200}, 2025{\natexlab{b}}.

\bibitem[Woo et~al.(2023)Woo, Debnath, Hu, Chen, Liu, Kweon, and
  Xie]{woo2023convnext}
Sanghyun Woo, Shoubhik Debnath, Ronghang Hu, Xinlei Chen, Zhuang Liu, In~So
  Kweon, and Saining Xie.
\newblock {ConvNeXt V2}: Co-designing and scaling {ConvNets} with masked
  autoencoders.
\newblock In \emph{Proceedings of the IEEE/CVF conference on computer vision
  and pattern recognition}, pp.\  16133--16142, 2023.

\bibitem[Xie et~al.(2025)Xie, Shen, Li, Xie, Tang, and
  Hu]{xie2025fireredtts2longconversationalspeech}
Kun Xie, Feiyu Shen, Junjie Li, Fenglong Xie, Xu~Tang, and Yao Hu.
\newblock Fireredtts-2: Towards long conversational speech generation for
  podcast and chatbot, 2025.
\newblock URL \url{https://arxiv.org/abs/2509.02020}.

\bibitem[Xu et~al.(2025{\natexlab{a}})Xu, Guo, He, Hu, He, Bai, Chen, Wang,
  Fan, Dang, Zhang, Wang, Chu, and Lin]{xu2025qwen25omnitechnicalreport}
Jin Xu, Zhifang Guo, Jinzheng He, Hangrui Hu, Ting He, Shuai Bai, Keqin Chen,
  Jialin Wang, Yang Fan, Kai Dang, Bin Zhang, Xiong Wang, Yunfei Chu, and
  Junyang Lin.
\newblock Qwen2.5-omni technical report, 2025{\natexlab{a}}.
\newblock URL \url{https://arxiv.org/abs/2503.20215}.

\bibitem[Xu et~al.(2025{\natexlab{b}})Xu, Guo, Hu, Chu, Wang, He, Wang, Shi,
  He, Zhu, Lv, Wang, Guo, Wang, Ma, Zhang, Zhang, Hao, Guo, Yang, Zhang, Ma,
  Wei, Bai, Chen, Liu, Wang, Yang, Liu, Ren, Zheng, Men, Zhou, Yu, Yang, Yu,
  Zhou, and Lin]{xu2025qwen3omnitechnicalreport}
Jin Xu, Zhifang Guo, Hangrui Hu, Yunfei Chu, Xiong Wang, Jinzheng He, Yuxuan
  Wang, Xian Shi, Ting He, Xinfa Zhu, Yuanjun Lv, Yongqi Wang, Dake Guo,
  He~Wang, Linhan Ma, Pei Zhang, Xinyu Zhang, Hongkun Hao, Zishan Guo, Baosong
  Yang, Bin Zhang, Ziyang Ma, Xipin Wei, Shuai Bai, Keqin Chen, Xuejing Liu,
  Peng Wang, Mingkun Yang, Dayiheng Liu, Xingzhang Ren, Bo~Zheng, Rui Men, Fan
  Zhou, Bowen Yu, Jianxin Yang, Le~Yu, Jingren Zhou, and Junyang Lin.
\newblock Qwen3-omni technical report, 2025{\natexlab{b}}.
\newblock URL \url{https://arxiv.org/abs/2509.17765}.

\bibitem[Yang et~al.(2025)Yang, Song, Zhuo, Cui, Li, Yang, Du, Ma, Liu, Wang,
  et~al.]{yang2025gigaspeech}
Yifan Yang, Zheshu Song, Jianheng Zhuo, Mingyu Cui, Jinpeng Li, Bo~Yang, Yexing
  Du, Ziyang Ma, Xunying Liu, Ziyuan Wang, et~al.
\newblock Gigaspeech 2: An evolving, large-scale and multi-domain asr corpus
  for low-resource languages with automated crawling, transcription and
  refinement.
\newblock In \emph{Proceedings of the 63rd Annual Meeting of the Association
  for Computational Linguistics (Volume 1: Long Papers)}, pp.\  2673--2686,
  2025.

\bibitem[Yao et~al.(2024)Yao, Yu, Zhang, Wang, Cui, Zhu, Cai, Li, Zhao, He,
  et~al.]{yao2024minicpm}
Yuan Yao, Tianyu Yu, Ao~Zhang, Chongyi Wang, Junbo Cui, Hongji Zhu, Tianchi
  Cai, Haoyu Li, Weilin Zhao, Zhihui He, et~al.
\newblock Minicpm-v: A gpt-4v level mllm on your phone.
\newblock \emph{arXiv preprint arXiv:2408.01800}, 2024.

\bibitem[Zeghidour et~al.(2025)Zeghidour, Kharitonov, Orsini, Volhejn,
  de~Marmiesse, Grave, P{'e}rez, Mazar{'e}, and
  D{'e}fossez]{zeghidour2025streaming}
Neil Zeghidour, Eugene Kharitonov, Manu Orsini, V{'a}clav Volhejn, Gabriel
  de~Marmiesse, Edouard Grave, Patrick P{'e}rez, Laurent Mazar{'e}, and
  Alexandre D{'e}fossez.
\newblock Streaming sequence-to-sequence learning with delayed streams
  modeling.
\newblock \emph{arXiv preprint arXiv:2509.08753}, 2025.

\bibitem[Zhang et~al.(2025)Zhang, Guo, Yang, Yu, Zhang, Lei, Mai, Yan, Yang,
  Yang, Huang, Jin, Jiang, Cheng, Li, Xiao, Zhou, Zhang, Lu, and
  He]{zhang2025minimaxspeechintrinsiczeroshottexttospeech}
Bowen Zhang, Congchao Guo, Geng Yang, Hang Yu, Haozhe Zhang, Heidi Lei, Jialong
  Mai, Junjie Yan, Kaiyue Yang, Mingqi Yang, Peikai Huang, Ruiyang Jin, Sitan
  Jiang, Weihua Cheng, Yawei Li, Yichen Xiao, Yiying Zhou, Yongmao Zhang, Yuan
  Lu, and Yucen He.
\newblock Minimax-speech: Intrinsic zero-shot text-to-speech with a learnable
  speaker encoder, 2025.
\newblock URL \url{https://arxiv.org/abs/2505.07916}.

\bibitem[Zheng et~al.(2024)Zheng, Yin, Xie, Sun, Huang, Yu, Cao, Kozyrakis,
  Stoica, Gonzalez, Barrett, and
  Sheng]{zheng2024sglangefficientexecutionstructured}
Lianmin Zheng, Liangsheng Yin, Zhiqiang Xie, Chuyue Sun, Jeff Huang, Cody~Hao
  Yu, Shiyi Cao, Christos Kozyrakis, Ion Stoica, Joseph~E. Gonzalez, Clark
  Barrett, and Ying Sheng.
\newblock Sglang: Efficient execution of structured language model programs,
  2024.
\newblock URL \url{https://arxiv.org/abs/2312.07104}.

\end{thebibliography}
\bibliographystyle{colm2024_conference}

\appendix

\section{Fish Audio Instruction Benchmark}\label{app:benchmark}

The evaluation of TTS models primarily relies on WER and MOS to measure speech intelligibility and timbre similarity.
However, as foundational speech models advance toward realistic human dialogue, these coarse-grained metrics fall short of evaluating the models' capability boundaries in fine-grained emotional control and paralinguistic expression.
Real human speech is rich in subtle physiological and emotional cues, including breaths, laughs, sighs, and emphasis.
To systematically quantify this capability, we propose the Fish Audio Instruction Benchmark, which aims to evaluate the zero-shot instruction-following abilities of models regarding inline vocol feature tags.

\subsection{Benchmark Design}
\label{subsec:benchmark_design}

In contrast to traditional datasets that rely on global emotion prompts (e.g., ``read in a sad tone''), our benchmark embeds precise, word-level acoustic event tags directly into the transcriptions. The dataset is constructed from utterance segments extracted from multi-turn dialogues and monologues.
Each sample retains the necessary contextual information and annotates designated text positions with target acoustic-event tags, such as \texttt{[laugh]}, \texttt{[whispers]}, \texttt{[inhale]}, \texttt{[exhale]}, and \texttt{[emphasis]}.
Models are required to infer these events from the semantic and conversational context and realize them at the corresponding positions during speech generation.
A representative test example in this format is shown in Figure~\ref{fig:test_sample}.

We construct our benchmark using two datasets, MELD~\citep{poria2019meld} and the game character voice dataset, to represent the English and Chinese settings, respectively.
MELD is a speech emotion dialogue corpus curated from the American TV series \emph{Friends}, while the Chinese subset is drawn from a community-curated corpus of game character voice lines.
For MELD, we randomly sample dialogues from the test split with lengths near the median, so as to avoid extremely short or overly long conversations.
For the chinese dataset, we randomly sample utterances from character story dialogues. Both subsets contain approximately 500 utterances.
We then apply the data pipeline described in Section~\ref{sec:data} to obtain coarse annotations, followed by expert human verification and fine-grained refinement to filter out audio noise, correct alignment errors, and ensure that inline tags are placed precisely and are semantically appropriate.

\subsection{Evaluation Method}

We adopt Gemini 3 Pro as an LLM-as-a-Judge to systematically evaluate fine-grained instruction-following ability from both local and global perspectives.
Specifically, we define three metrics:
\begin{itemize}[leftmargin=2em, itemsep=-0.3em, topsep=-0.2em]
  \item \textbf{Tag Activation Rate (TAR)}: the proportion of inline tags (e.g., [laugh], [inhale]) for which the model successfully triggers the target event at the exact specified position, reported as a percentage.
  \item \textbf{Acoustic Naturalness} (1--5): a subjective realism metric that assesses whether the triggered vocal events sound natural and human-like, or exhibit noticeable synthetic artifacts (e.g., robotic timbre, distortion, or abrupt insertions).
  \item \textbf{Global Expressiveness} (1--5): a holistic expressiveness metric that evaluates whether the overall prosody and affective tone of an utterance align with the semantic content and contextual cues implied by the text and tags (e.g., utterances containing \texttt{[whispers]} should not be delivered in a broadcast-like style or with overly projected voice).
\end{itemize}

The evaluation procedure consists of two steps.
We first synthesize speech for each benchmark sample using the target TTS model; we then feed the generated audio, along with its corresponding text and inline tags, into Gemini 3 Pro to obtain decisions/scores for the three dimensions above.
Finally, we aggregate the results across all samples to quantify overall performance under different languages and scenarios.

\subsection{Human-Model Alignment Validation}

Before using Gemini 3 Pro for large-scale automated evaluation, we assess whether its scoring behavior is aligned with human auditory judgment.
We perform stratified random sampling and select 200 audio clips with diverse acoustic-event tags from baseline model outputs.
A professional annotation team then evaluates these clips under double-blind conditions using the same rubric as the Gemini 3 Pro.

We quantify agreement for both objective event detection and subjective perceptual ratings.
For objective acoustic-event detection, the absolute agreement accuracy between Gemini 3 Pro and human experts reaches $76.2\%$, with Cohen's kappa ($\kappa$) of $0.47$.
This indicates moderate agreement, suggesting that the model provides a reasonable approximation of human judgment for basic event detection.
For 1--5 Likert-scale subjective ratings, Pearson correlations ($r$) for Naturalness and Expressiveness are $0.55$ and $0.42$, while the corresponding Quadratic Weighted Kappa (QWK) scores are $0.36$ and $0.47$.
Given the inherent subjectivity of expressive-speech assessment, these results indicate a clear positive alignment trend between model judgments and human perception.
Although a gap remains in fine-grained absolute scoring, the observed ranking consistency supports the use of Gemini 3 Pro as a practical tool for large-scale preliminary screening and relative quality assessment.

\subsection{Conclusion, limitations and future work}

The Fish Audio Instruction Benchmark addresses an important gap in TTS evaluation by providing a dedicated testbed for fine-grained instruction following. As a first release, it still has limitations, including limited data diversity, imbalanced acoustic-tag distributions, and early-stage human--model alignment analysis. In future versions, we will expand the dataset and improve the automated evaluation pipeline to provide a more comprehensive and reliable standard.

\end{document}